\documentclass[aps, preprint, groupedaddress, floatfix]{revtex4}
\usepackage{epsfig}
\usepackage{graphicx}
\usepackage{amsmath}
\usepackage{bbold}
\usepackage{tabularx, booktabs}
\usepackage[normalem]{ulem}
\usepackage{xcolor}

\newcommand{\etal}{{\it{et al.}}}
\newcommand{\ie}{{\it i.e.}}

\newcolumntype{Y}{>{\centering\arraybackslash}X}
\newcolumntype{Z}{>{\hsize=1.1\hsize\centering\arraybackslash}X}

\newcommand*{\SSRO}{Sr$_2$ScRuO$_6$}
\newcommand*{\SSOO}{Sr$_2$ScOsO$_6$}
\newcommand*{\SYOO}{Sr$_2$YOsO$_6$}
\newcommand*{\SYRO}{Sr$_2$YRuO$_6$}
\newcommand*{\BYRO}{Ba$_2$YRuO$_6$}
\newcommand*{\BYOO}{Ba$_2$YOsO$_6$}

\begin{document}

\title{
The complex non-collinear magnetic orderings in \BYOO: A new approach to tuning spin-lattice interactions and controlling magnetic orderings in frustrated complex oxides
}

\author{Yue-Wen Fang$^{1,2}$}
\email{fyuewen@gmail.com}
\author{Ruihan Yang$^{2,3}$}
\author{Hanghui Chen$^{2,4}$}
\email{hanghui.chen@nyu.edu}
\affiliation{$^1$Department of Materials Science and Engineering, Kyoto University, Kyoto 606-8501, Japan \\
$^2$NYU-ECNU Institute of Physics, New York University Shanghai China \\
$^3$Department of Engineering and Computer Science, New York University Shanghai China \\
$^4$Department of Physics, New York University, New York  10003, USA
}


\begin{abstract}
Frustrated magnets are one class of fascinating materials that host
many intriguing phases such as spin ice, spin liquid and complex
long-range magnetic orderings at low temperatures.  In this work we
use first-principles calculations to find that in a wide range of
magnetically frustrated oxides, at zero temperature a number of
non-collinear magnetic orderings are more stable than the type-I
collinear ordering that is observed at finite temperatures. The
emergence of non-collinear orderings in those complex oxides is due to
higher-order exchange interactions that originate from second-row
and third-row transition metal elements. This implies a
collinear-to-noncollinear spin transition at sufficiently low
temperatures in those frustrated complex oxides. Furthermore, we find
that in a particular oxide Ba$_2$YOsO$_6$, experimentally feasible
uniaxial strain can tune the material between two different
non-collinear magnetic orderings. Our work predicts new non-collinear
magnetic orderings in frustrated complex oxides at very low
temperatures and provides a mechanical route to tuning complex
non-collinear magnetic orderings in those materials.
\end{abstract}
\maketitle

\section{Introduction}

Magnetic frustration, arising either from the geometry of crystal
lattice or from the competition between different magnetic
interactions, can lead to many intriguing phenomena such as
complex long-range ordered states (non-collinear, chiral, etc.) and
disordered states (spin liquid, spin ice,
etc.)~\cite{balz-natphy-2016VB,li-natcomm-2017valencebond,Bojesen-PRL-spinice2017,PRX-spinice-2018,balents-nature-2010spinliquid,rousochatzakis-natcomm-spinliquid2018,savary2016Leon-review,zhang2017quantum,RevModPhys.88.041002}. However,
while frustrated magnetism is extensively studied in model
calculations and experiments, first-principles studies on
realistic frustrated magnetic materials are few, compared to those on
normal magnets with a bipartite lattice and a dominating magnetic
interaction. In particular, insights from first-principles
  calculations on low-temperature magnetism in frustrated complex
  oxides are rare~\cite{kayser2017spin-organicchem,PhysRevB.91.100406,PhysRevB.93.220408,PhysRevLett.110.017202,PhysRevB.91.054415,ZAAC:ZAAC201400590,PhysRevB.98.104434}.
  Density-functional-theory-based first-principles study can take into
  account various exchange interactions, spin-orbit interaction and
  spin-lattice interactions in realistic materials and treat them on
  an equal footing, enabling us to systematically search for new
  magnetic phenomena in frustrated complex materials.

In this work, we study a wide range of ordered double perovskite
  oxides with second-row or third-row transition metal elements residing
  on a geometrically frustrated face-centered-cubic lattice: \SSRO,
  \SYRO, \BYRO, \SSOO, \SYOO, and \BYOO. All these complex oxides are
  all reported in experiment to exhibit type-I collinear magnetic
  ordering below the respective N\'{e}el
  temperatures~\cite{doi:10.1021/acs.inorgchem.7b00983,BATTLE1984138,IZUMIYAMA2002125,PhysRevB.91.100406,ZAAC:ZAAC201400590,Kermarrec-BYOO-PRB2015}.
  Our first-principles calculations show that at zero temperature, a
  number of non-collinear magnetic orderings are more stable than the
  type-I collinear magnetic ordering that is observed at finite
  temperatures. The emergence of non-collinear orderings in those
complex oxides at low temperatures is due to higher-order exchange
interactions that originate from second-row and third-row transition
metal elements.
This implies that at sufficiently low temperatures
there could occur a collinear-to-noncollinear spin transition in those
frustrated complex oxides. Furthermore, we find that in a particular
oxide Ba$_2$YOsO$_6$, experimentally feasible uniaxial strain can tune
the material between two different non-collinear magnetic orderings,
which reveals a mechanical approach to tuning spin-lattice
interactions and controlling magnetic ordering in frustrated magnets.
Our conclusions are robust against different exchange-correlation
functionals, correlation strength of magnetic ions and whether
spin-orbit coupling is taken into account or not.

\begin{figure}[t]
\includegraphics[angle=0,width=0.8\textwidth]{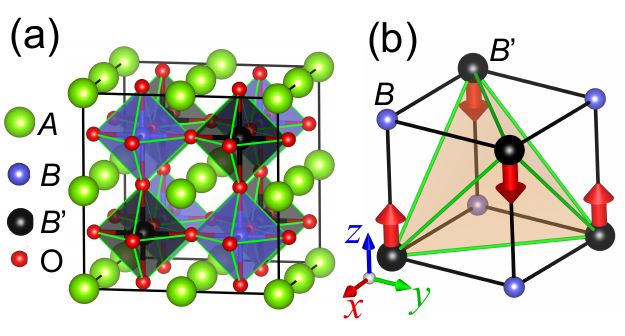}
\caption{\label{fig:fig1} \textbf{a}) A complete crystal structure of
  ordered double perovskite oxide ${A_{2}BB'}$O$_6$: the green, blue,
  black and red balls correspond to $A$-site ions, $B$-site
  non-magnetic ions, $B'$-site magnetic ions and O.  \textbf{b}) A
  simplified crystal structure with only $B$-site (blue) and $B'$-site
  (black) ions are shown.  The red bold arrows indicate spins. The
  $B'$-site magnetic ions occupy a frustrated face-centered-cubic
  lattice, which is highlighted by the shallow orange tetrahedron.
  }
\end{figure}

The frustrated complex oxides in this study have a double perovskite
structure with a chemical formula $A_2BB'$O$_6$ (see
figure~\ref{fig:fig1}(a)). Two different types of transition metal
ions $B$ and $B'$ form a rock-salt ordering and therefore transition
metal ions of the same type ($B$ or $B'$) occupy a face-centered-cubic
lattice (fcc), which has `geometric frustration'.  We study six double
perovskite oxides which are listed in table~\ref{tab:material}. They
all have one non-magnetic ion (Sc$^{3+}$ and Y$^{3+}$) and one
magnetic ion (Ru$^{5+}$ and Os$^{5+}$).  Experimentally, all these six
complex oxides exhibit a layered collinear antiferromagnetic ordering
(so-called type-I) at finite temperatures, as is shown in
figure~\ref{fig:fig1}(b).
The N\'{e}el temperatures of these six
complex oxides are listed in table~\ref{tab:material}.

\begin{table}[t]
\caption{A list of magnetically frustrated complex oxides in this
  study. FM and AFM-I refer to ferromagnetic ordering and type-I
  antiferromagnetic ordering, respectively.  Note that the notations
  of space groups in this table are taken from the respective
  references directly.}
\label{tab:material}
\begin{tabularx}{0.85\textwidth}{cccl l l l l l} 
 \hline \hline 
Material  & Magnetic ion & $d$ Shell  & Space group  & Magnetic transition & Ref.\\
		\hline
\SSRO & Ru$^{5+}$ & 4$d^{3}$ & ${I2/m}$ (300 K)   &  AFM-I, $T_{\rm N}\sim$~60 K  & \cite{doi:10.1021/acs.inorgchem.7b00983} \\
\SYRO & Ru$^{5+}$ & 4$d^{3}$ & ${P2_{1}/n}$ (293 K)   &  AFM-I, $T_{\rm N}\sim$~26 K   & \cite{BATTLE1984138} \\ 
\BYRO & Ru$^{5+}$  &  4$d^{3}$ & ${Fm\bar{3}m}$ (4.2 K)   & AFM-I, $T_{\rm N}\sim$~37 K  & \cite{BATTLE1989108,IZUMIYAMA2002125} \\
\SSOO & Os$^{5+}$ &  5$d^{3}$ & ${P2_{1}/n}$ (3.5 to 300 K)  &   AFM-I, $T_{\rm N}\sim$~92 K   &  \cite{PhysRevB.91.100406}\\
\SYOO & Os$^{5+}$ &  5$d^{3}$ & ${P2_{1}/n}$ (2.9 to 300 K)  &   AFM-I, $T_{\rm N}\sim$~53 K  & \cite{ZAAC:ZAAC201400590} \\
\BYOO & Os$^{5+}$ &  5$d^{3}$ & ${Fm\bar{3}m}$ (3.5 to 290 K)   &  AFM-I, $T_{\rm N}\sim$~69 K &  \cite{Kermarrec-BYOO-PRB2015}\\
\hline
\hline
\end{tabularx}
\end{table}

\section{Computational details}
\label{Section_Methods}

We perform first-principles calculations using plane-wave basis
density functional theory (DFT), as implemented in the Vienna
Ab-initio Simulation Package
(VASP)~\cite{Kresse1996,Kresse-PRB-1996}. We take into account both
spin-orbit coupling (SOC) and correlation effects.  We use the method
proposed by Dudarev~\etal~\cite{Dudarev-LDAU-PRB1998} to model Hubbard
$U$ interaction. We employ a revised Perdew-Burke-Ernzerhof
generalized gradient approximation (PBEsol)
~\cite{PhysRevLett.100.136406PBEsol} as the exchange-correlation
functional, which has been successfully applied to study second-row
and third-row transition metal
oxides~\cite{Chen_npj2018,aulesti2018APL}.  We also test other
exchange-correlation functionals: local-density-approximation (LDA)
and Perdew-Burke-Ernzerhof functional (PBE). All the calculations are
spin-polarized (with either collinear and non-collinear magnetic
ordering). We use an energy cutoff of 600 eV, and the Brillouin zone
integration is performed with a Gaussian smearing of 0.05 eV and a 10
$\times$ 10 $\times$ 10 \textbf{k}-mesh.  The threshold of energy
convergence is $10^{-6}$ eV. Throughout the calculations we use a
40-atom supercell (see figure~\ref{fig:fig1}(a)).  For each oxide, we use
its experimental structure to compare different long-range magnetic
orderings.  Their experimental crystal structures can be found in the
references listed in table~\ref{tab:material}.
Only when we study uniaxial strain,
we relax the crystal structure of Ba$_2$YOsO$_6$ until each
Hellmann-Feynman force component is smaller than $10^{-3}$ eV/\AA~and
the stress tensor is smaller than 1 kbar.

\section{Results and discussion}

For clarity, we first study Ba$_2$YOsO$_6$ as a representative example
and then extend the discussion to other complex oxides.
Experimentally double perovskite Ba$_2$YOsO$_6$ crystallizes in a
cubic structure with a lattice constant of 8.357~\AA~
(space group $Fm\bar{3}m$)~\cite{BYOO-expstructure-1981-german}.
In Ba$_2$YOsO$_6$, Y$^{3+}$ nominally has a $d^0$ occupancy and
Os$^{5+}$ nominally has a $d^3$ occupancy. Due to Hund's rule, the
three electrons on Os$^{5+}$ ions fill three different $t_{2g}$
orbitals and form a core spin $S =
3/2$~\cite{ChenGang_PRB2011dp-d2}. The Os spins occupy a
face-centered-cubic lattice, which has `geometric frustration' (see
figure~\ref{fig:fig1}(b)).
The experimental Os$^{5+}$ moment is
1.65$\mu_B$~\cite{Kermarrec-BYOO-PRB2015}, which is smaller than the
atomic value 3$\mu_B$ for a $S = 3/2$ spin, due to strong
hybridization of Os-$5d$ orbitals with O-$2p$ orbitals. 

We discuss our results in two steps: in the first step we do not
consider spin-orbit coupling (SOC) and in the second step we
take into account SOC effects. This is to decouple the
SOC effects from other intrinsic spin interactions.

\subsection{Without spin-orbit coupling}

\begin{figure}[t]
\includegraphics[angle=0,width=0.5\textwidth]{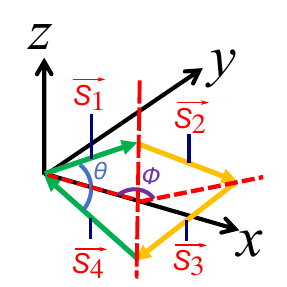}
\caption{\label{fig:fig2-JPCM} A general 4-sublattice
  antiferromagnetic spin configuration in three-dimensional space. The
  four spins (equal-length vector) form a head-to-tail ring with the
  two green spins in one plane and the other two yellow spins in
  another plane. The ring formed by the four spins is characterized by
  two angles $\theta$ and $\phi$. Here, $\theta$ is the angle between
  the two green spins, and $\phi$ is the angle between the two
  planes.}
\end{figure}

Without taking into account SOC, spins are decoupled to lattice.  This
means that the total energy of the system only depends on relative
orientations between spins. Any global rotation of the full spin
configuration with respect to lattice leads to a trivial degenerate
state.

Since our 40-atom supercell includes four Os atoms that occupy a tetrahedron,
we study a general 4-sublattice antiferromagnetic ordering, as well as
a ferromagnetic ordering for comparison.  The spin configuration of a
general 4-sublattice antiferromagnetic ordering is schematically shown
in figure~\ref{fig:fig2-JPCM} in which four equal-length arrows (\ie,
vector spins) form a head-to-tail ring. In three-dimensional space,
such a ring is characterized only by two parameters $\theta$ and
$\phi$, as figure~\ref{fig:fig2-JPCM} shows. Thus, the four spins
have the following coordinates:

\begin{eqnarray}
\label{eq1} 
\textbf{S}_1 && = \left(+\cos\left(\theta/2\right),
  0, +\sin\left(\theta/2\right)\right) \\ \nonumber
\textbf{S}_2 && = \left(+\cos\left(\theta/2\right)\cos(\pi-\phi),
  +\cos(\theta/2)\sin(\pi-\phi), -\sin\left(\theta/2\right)\right) \\ \nonumber
\textbf{S}_3 && =  \left(-\cos\left(\theta/2\right)\cos(\pi-\phi),
  -\cos(\theta/2)\sin(\pi-\phi), -\sin\left(\theta/2\right)\right) \\ \nonumber
\textbf{S}_4 &&=   \left(-\cos\left(\theta/2\right),
  0, +\sin\left(\theta/2\right)\right)
\end{eqnarray}

\begin{figure}[t]
\includegraphics[angle=0,width=0.89\textwidth]{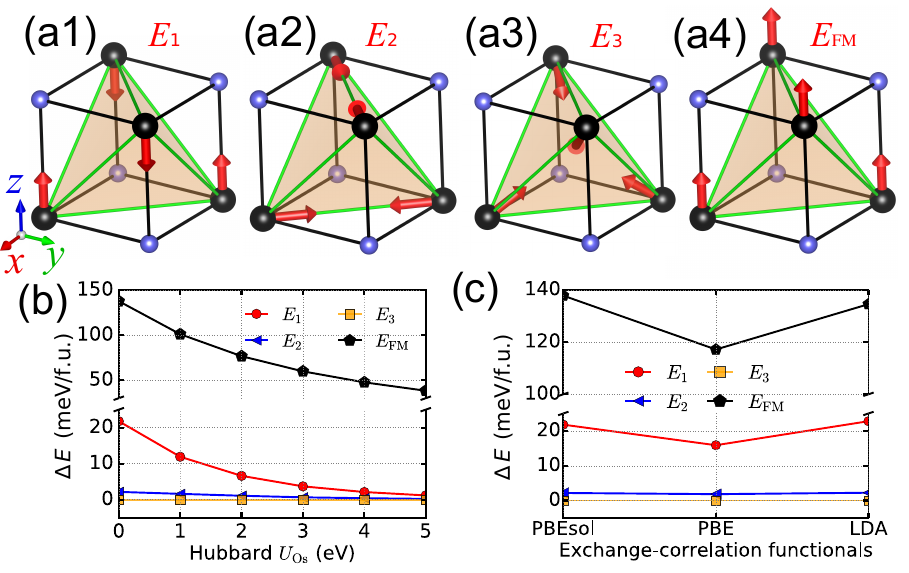}
\caption{\label{fig:fig2} The magnetic orderings that are stabilized
  in Ba$_2$YOsO$_6$ from spin-polarized DFT-PBEsol calculations
  without spin-orbit coupling (SOC).  \textbf{a1}) Collinear
  antiferromagnetic ordering, referred to as $E_1$. \textbf{a2})
  Coplanar antiferromagnetic ordering, referred to as $E_2$.
  \textbf{a3}) Non-collinear non-coplanar antiferromagnetic ordering
  with all four spins pointing towards the center of the tetrahedron,
  referred to as $E_3$. \textbf{a4}) Ferromagnetic ordering, referred
  to as $E_{\textrm{FM}}$.  \textbf{b}) Total energies of collinear
  state ($E_1$), coplanar state ($E_2$) and non-collinear non-coplanar
  state ($E_3$) as well as ferromagnetic state ($E_{\textrm{FM}}$) of
  Ba$_2$YOsO$_6$ as a function of Hubbard $U_{\textrm{Os}}$ in
  spin-polarized DFT-PBEsol+$U$ calculations (without taking into account SOC).
  \textbf{c}) Total energies of the four magnetic orderings
  ($E_1$, $E_2$, $E_3$ and $E_{\textrm{FM}}$) calculated by different
  exchange-correlation functionals (without taking into account SOC).
  The energy of non-collinear non-coplanar state ($E_3$) is chosen as the zero
  point. Note that panels \textbf{b}) and \textbf{c}) use broken
  energy axes.}
\end{figure}

We first perform spin-polarized DFT-PBEsol calculations using the
experimental structure of Ba$_2$YOsO$_6$ (see table S1 in section I of
Supplementary Materials for its structural parameters).
Then we
discuss $U$ dependence and exchange-correlation functional dependence.
We compute the total energy of the ferromagnetic ordering and
different 4-sublattice antiferromagnetic orderings (collinear,
coplanar, non-collinear non-coplanar etc.).  Our spin-polarized
DFT-PBEsol calculations find that three distinct antiferromagnetic
states as well as ferromagnetic ordering are stabilized in
Ba$_2$YOsO$_6$. They are shown in figure~\ref{fig:fig2}.
Figure~\ref{fig:fig2}(a1) is a collinear
antiferromagnetic state in which all the spins are either parallel or
anti-parallel ($\theta=180^{\circ}$, referred to as
$E_1$).
For simplicity, we use $E$ to refer to a state as well as the energy
of that state.
Figure~\ref{fig:fig2}(a2) is a coplanar
antiferromagnetic state in which all four spins lie in the same plane;
one pair of anti-parallel spins is orthogonal to another pair of
anti-parallel spins ($\theta=0^{\circ}, \phi=90^{\circ}$, referred to
as $E_2$). Figure~\ref{fig:fig2}(a3) is a non-collinear non-coplanar
state in which every two spins form an identical angle
($\theta=\arccos\left(\frac{1}{3}\right)\simeq 71^{\circ},
\phi=90^{\circ}$, referred to as $E_3$). Figure~\ref{fig:fig2}(a4) is
a ferromagnetic state (referred to as $E_{\textrm{FM}}$).
Figure~\ref{fig:fig2}(b) shows that the non-collinear non-coplanar
state ($E_3$) has the lowest total energy, followed by the coplanar
state ($E_2$) and then followed by the collinear state ($E_1$). The
ferromagnetic state ($E_{\textrm{FM}}$) has much higher energy than
all antiferromagnetic orderings, which indicates that the
nearest-neighbor exchange coupling is antiferromagnetic in nature.

\begin{figure}[t!]
\includegraphics[angle=0,width=0.8\textwidth]{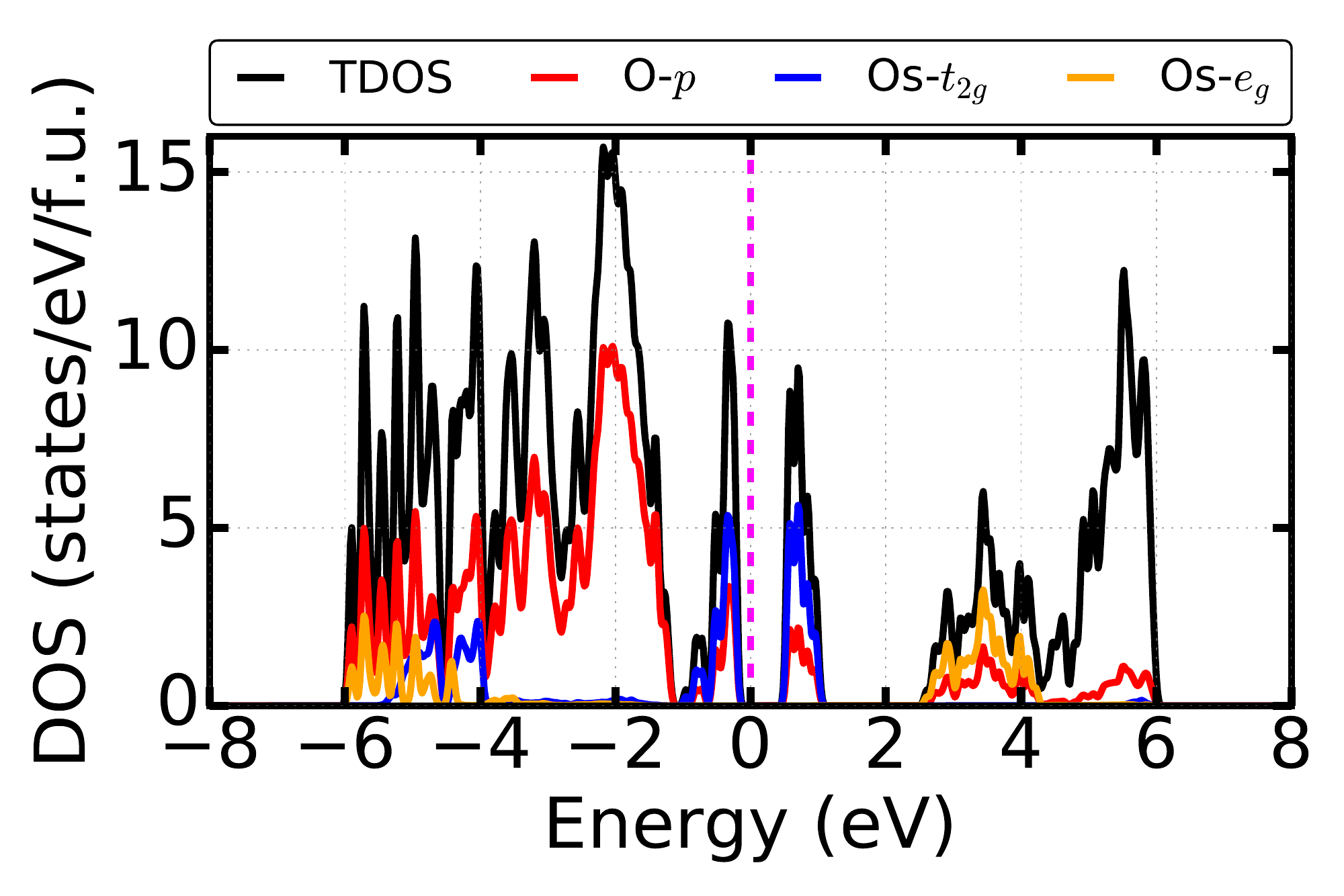}
\caption{\label{fig:dos} Total density of states and projected
  densities of states of Ba$_2$YOsO$_6$ in a non-collinear
  non-coplanar state ($E_3$), calculated using spin-polarized
  DFT-PBEsol method. The black curve is the total density of
  states. The red curve is the O-$p$ projected density of states. The
  blue and orange curves are Os-$t_{2g}$ and Os-$e_{g}$ projected
  densities of states, respectively.  The Fermi level is at zero energy, highlighted
  by the magenta dashed line.}
\end{figure}

Next, we discuss Hubbard $U$ dependence and exchange-correlation functional
dependence. The correlated ion in
Ba$_2$YOsO$_6$ is Os$^{5+}$. While the accurate value of Hubbard $U$
on Os is not known,
we expect that it does not exceed 5 eV because
Os is a third-row transition metal element~\cite{LAN2018909,PhysRevB.89.214414}.
We repeat the
previous calculations using different values of $U_{\textrm{Os}}$
ranging from 0 to 5 eV. The results are shown in
figure~\ref{fig:fig2}(b).  We find that while
the energy difference between the three antiferromagnetic
orderings decreases with $U_{\textrm{Os}}$,
the energy sequence $E_3 < E_2 < E_1 < E_{\textrm{FM}}$ does not change
with Hubbard $U_{\textrm{Os}}$. On the other hand, the magnitude of
Os-projected magnetic moment increases with $U$ (from $m_{\textrm{Os}}
= 1.8~{\mu_B}$ at $U_{\textrm{Os}}=0$ eV to $m_{\textrm{Os}} =
2.5~{\mu_B}$ at $U_{\textrm{Os}}=5$ eV). Experimentally
$m_{\textrm{Os}}= 1.65~{\mu_B}$~in
Ba$_2$YOsO$_6$~\cite{Kermarrec-BYOO-PRB2015} and the
$U_{\textrm{Os}}=0$ result is the closest to the experimental
value. This is consistent with previous studies showing that in
spin-polarized DFT calculations, the
exchange splitting in PBEsol exchange-correlation functional is
sufficiently large~\cite{JiaChen2015, HChen-2016-PRB}. Turning on a
Hubbard $U$ impairs the agreement between experiment and
theory. We therefore use $U_{\textrm{Os}}=0$ eV in the remainder of
the paper. In addition, we study the effect of different exchange-correlation
functionals on the energy sequence of the four magnetic orderings.  We
compare the energy differences ${\Delta E}$ obtained by spin-polarized
DFT calculations using PBEsol, PBE and LDA calculations.
Figure~\ref{fig:fig2}(c) shows that the energy sequence
$E_3 < E_2 < E_1 < E_{\textrm{FM}}$ does not change in all the calculations,
indicating that our results are robust.

\begin{figure}[t!]
\includegraphics[angle=0,width=0.8\textwidth]{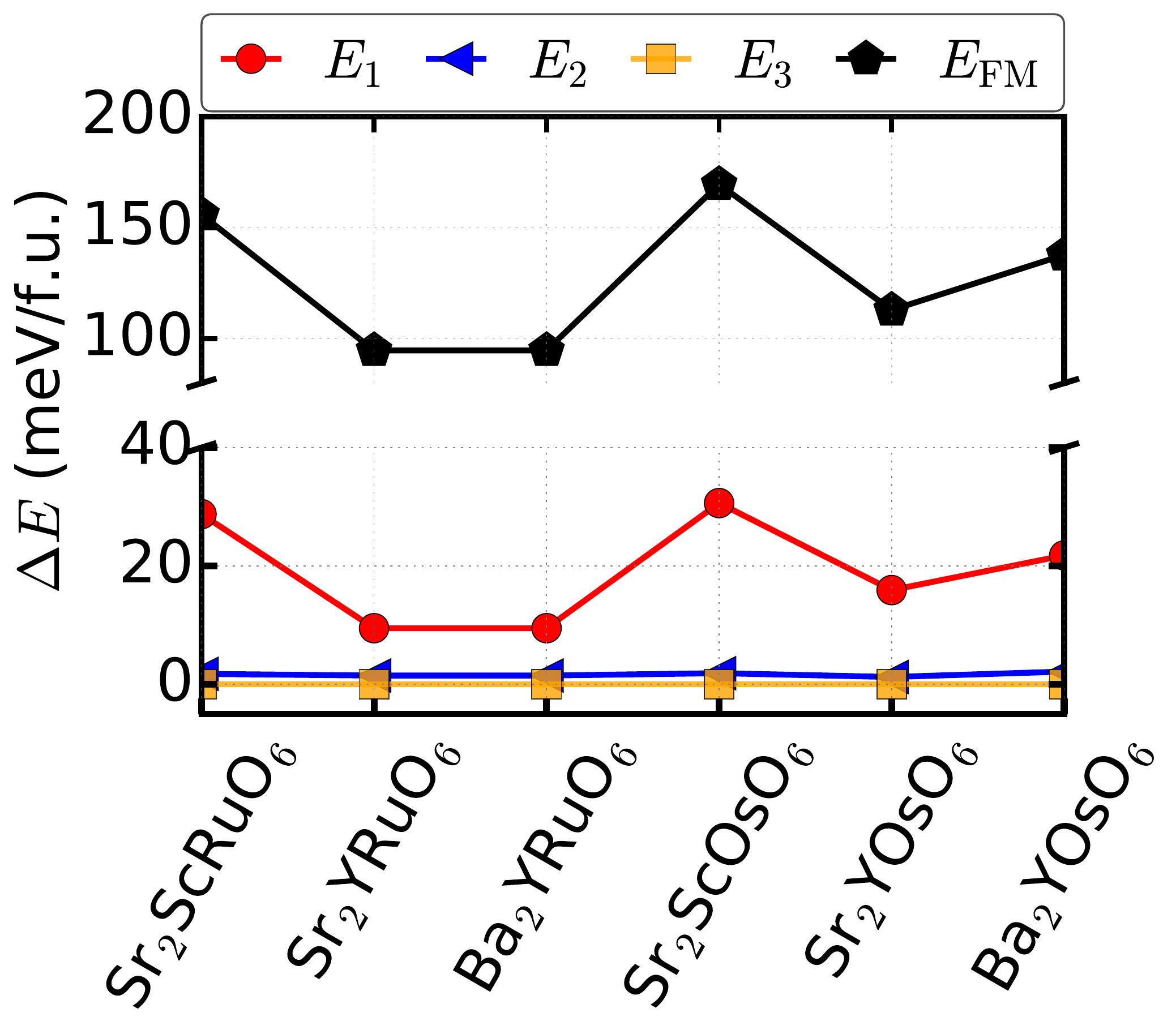}
\caption{\label{figs:validate-E-ordering} The total energies of
  ferromagnetic ordering ($E_{\rm FM}$), collinear antiferromagnetic
  ordering ($E_1$), coplanar antiferromagnetic ordering ($E_2$) and
  noncollinear-noncoplanar antiferromagnetic ordering ($E_3$) for all
  the six complex oxides (without taking into account SOC).  The state
  of non-collinear non-coplanar state ($E_3$) is chosen as the zero
  energy.  Note that we use a broken energy axis because the energy of
  ferromagnetic ordering is much higher than antiferromagnetic
  orderings.}
\end{figure}

Figure~\ref{fig:dos} shows the total density of states and
projected densities of states of Ba$_2$YOsO$_6$ in the lowest-energy
non-collinear non-coplanar magnetic ordering ($E_3$), calculated by
spin-polarized DFT-PBEsol method. The other magnetic orderings have
similar densities of states. Close to the Fermi level are dominating
Os-$t_{2g}$ states, while Os-$e_g$ states are much higher than the
Fermi level. Therefore, the low-energy physics can be well described by
Os-$t_{2g}$ states which form a $S = 3/2$ core spin.

Now we repeat the spin-polarized DFT-PBEsol calculations on the other
five complex oxides and find the same energy sequence $E_3 < E_2 < E_1
< E_{\textrm{FM}}$ between the three antiferromagnetic orderings and
the ferromagnetic ordering (see
figure~\ref{figs:validate-E-ordering}). Turning on Hubbard $U$ on Ru
or Os does not qualitatively change the results.  This implies that
the underlying spin interaction could be universal to this class of
frustrated magnets. Considering that the spins of Os$^{5+}$ and
Ru$^{5+}$ are large ($S=3/2$), we construct the spin interaction using
a classical vector-spin model.  The simplest nearest-neighbor
classical Heisenberg interaction predicts that all 4-sublattice
antiferromagnetic orderings on a fcc-lattice are continuously
degenerate~\cite{Oguchi1985}. This is clearly at odds with the
DFT results that all three stable antiferromagnetic orderings
have different energies. The second nearest-neighbor Heisenberg
interaction is trivial in our DFT calculations since we use a
40-atom supercell which includes four magnetic ions. For a given magnetic
ion, the other three magnetic ions in the cell are its
nearest-neighbors on a fcc lattice. Its second nearest-neighbors are
in fact the periodic images in the adjacent cell. Therefore, the second
nearest-neighbor interaction is a constant in our DFT calculations
because the two spins are identical.

A common beyond-Heisenberg spin interaction is the nearest-neighbor
biquadratic interactions and 4-spin ring interactions, which have been
shown to be important in complex oxides~\cite{Calzado2003, Coldea2001,
  Xiang2013, Fedorova2015}.  They may arise from spin-lattice
interaction~\cite{Penc2004, Penc2007, Shannon2010}, or have pure
electronic origin~\cite{Kaplan2009, Takahashi1977}.  The latter is
relevant to our case because compared to first-row transition metal
elements, second-row element Ru and third-row element Os have a
smaller interaction strength $U$ due to stronger
hybridization between transition metal elements and oxygen. They also have a
larger inter-site hopping matrix elements $t$ owing to more extended
$4d$ and $5d$ orbitals.  The two factors combined lead to more terms
in the $t/U$ expansion of a half-filled Hubbard model,
the leading-order term of which is the nearest-neighbor Heisenberg
interaction.  The next-order terms are nearest-neighbor biquadratic
interactions and 4-spin ring interactions. Therefore, a vector-spin
Hamiltonian can be written as:
\begin{eqnarray}
  \label{eq2} &H= H_{\textrm{Heisenberg}} + H_{\textrm{bi-qudratic}} + H_{\textrm{4-ring}} + H_0\\
  &=
\frac{J_1}{2}\sum_{\langle ij \rangle}\textbf{S}_i\cdot\textbf{S}_j  +
\frac{a_1}{2}\sum_{\langle ij
  \rangle}(\textbf{S}_i\cdot\textbf{S}_j)^2+\frac{a_2}{2}\sum_{\langle
  ijkl \rangle}(\textbf{S}_i\cdot\textbf{S}_j)(\textbf{S}_k\cdot\textbf{S}_l)
+ NE_0\nonumber
\end{eqnarray} 
where $|\textbf{S}_i|=1$ is a vector-spin, $N$ is the number of spins
and $E_0$ is a reference energy. First we consider a general 4-sublattice
antiferromagnetic ordering on a fcc lattice and insert Eq.~(\ref{eq1})
into Eq.~(\ref{eq2}), we obtain:
\begin{equation}
\label{eq3}\frac{E}{N}
=-2J_1+(a_1+a_2)\left(\frac{13}{4}-\cos\theta+\frac{7}{4}\cos(2\theta)+2\cos\left(\frac{\theta}{2}\right)^4\cos(2\phi)\right) + E_0
\end{equation}
The nearest-neighbor Heisenberg interaction ($J_1$-term) does not
depend on $\theta$ and $\phi$, indicating a continuous degeneracy, as
we mentioned above. The nearest-neighbor biquadratic and 4-ring interactions
are additive on a fcc lattice, indicating that we can combine
the two interactions with one coefficient $\alpha_1=a_1+a_2$.
Eq.~(\ref{eq3}) have three extremal
solutions, which exactly correspond to the collinear
state ($E_1$), the coplanar state ($E_2$) and the non-collinear
non-coplanar state ($E_3$). Their total energies are:
\begin{eqnarray}
\label{eq23XXX} E_1/N = -2J_1 + 6 \alpha_1  +E_0 \\ \nonumber
E_2/N = -2 J_1 + 2 \alpha_1  +E_0 \\ \nonumber
E_3/N = -2 J_1 + \frac{2}{3} \alpha_1 +E_0
\end{eqnarray}
We show the details of derivating three extremes in section II of Supplementary Materials.
Next, we consider ferromagnetic ordering in the model Eq.~(\ref{eq2}) and
it is easy to get:
\begin{equation}
\label{eqFM}E_{\textrm{FM}}/N = 6J_1 + 6 \alpha_1  +E_0
\end{equation}
For a positive $J_1$ and a positive $\alpha_1$, Eq.~(\ref{eq23XXX}) and
Eq.~(\ref{eqFM}) find an energy
sequence $E_3 < E_2 < E_1 < E_{\textrm{FM}}$, irrespective of the values
of $J_1$ and $\alpha_1$. This reproduces our spin-polarized DFT results
(figure~\ref{figs:validate-E-ordering}).

We note that Hubbard $U$ on Ru or Os atoms can change the magnitude of
the coefficients $J_1$ and $\alpha$ by $J_1 \propto t^2/U$ and
$\alpha_1 \propto t^3/U^2$~\cite{Fedorova2015}. However, Hubbard $U$
can \textit{not} change the sign of $J_1$ and $\alpha$. A
positive $J_1$ and a positive $\alpha_1$ always lead to
the energy sequence $E_3 < E_2 < E_1 < E_{\textrm{FM}}$. 
On the other hand, as we
increase the Hubbard $U$ on Ru or Os, the energy difference
between the four magnetic orderings gets smaller as Eq.~(\ref{eq23XXX})
and Eq.~(\ref{eqFM}) indicate.
This is consistent with
the spin-polarized DFT+$U$ results shown in figure~\ref{fig:fig2}(b).

\subsection{With spin-orbit coupling}

\begin{figure}[ht!]
\includegraphics[angle=0,width=0.9\textwidth]{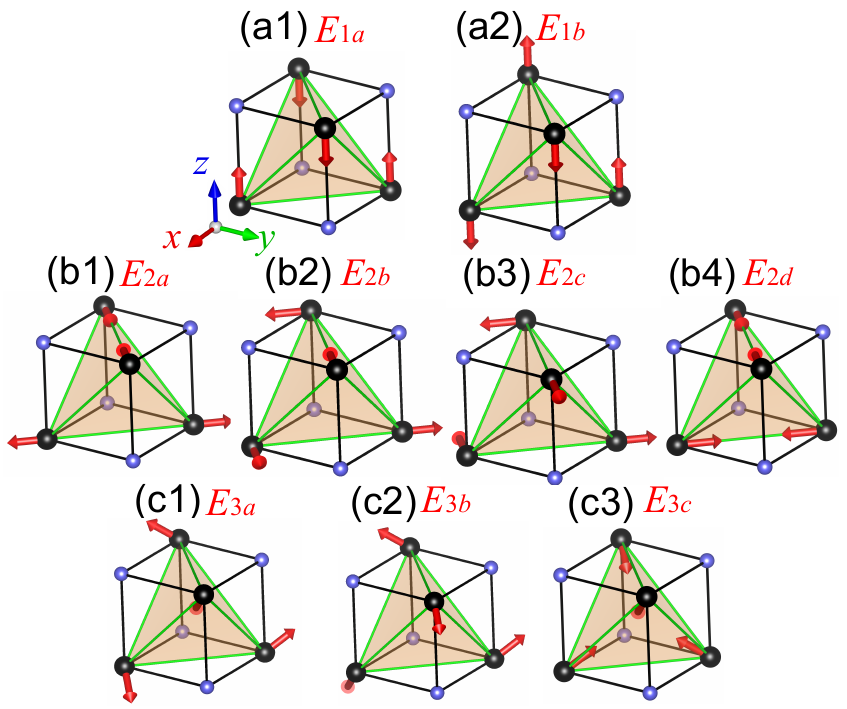}
\caption{\label{fig:fig3} Nine stable antiferromagnetic orderings
  found in Ba$_2$YOsO$_6$ from spin-polarized DFT-PBEsol
  calculations with spin-orbit
  coupling (PBEsol+SOC).  \textbf{a}) Two collinear antiferromagnetic
  orderings: \textbf{a1}) referred to as $E_{1a}$; \textbf{a2})
  referred to as $E_{1b}$.  \textbf{b}) Four coplanar
  antiferromagnetic orderings: \textbf{b1}) referred to as $E_{2a}$;
  \textbf{b2}) referred to as $E_{2b}$; \textbf{b3}) referred to as
  $E_{2c}$; \textbf{b4}) referred to as $E_{2d}$.  \textbf{c}) Three
  non-collinear non-coplanar antiferromagnetic orderings: \textbf{c1})
  referred to as $E_{3a}$; \textbf{c2}) referred to as $E_{3b}$;
  \textbf{c3}) referred to as $E_{3c}$.}
\end{figure}

Ru and Os are heavy elements and their SOC cannot be neglected. We
perform spin-polarized DFT-PBEsol+SOC calculations to study SOC
effects on the zero-temperature magnetic orderings in those complex
oxides. Similarly we first study Ba$_2$YOsO$_6$ as a representative
example and then extend the discussion to the other five oxides.

\begin{figure}[t!]
\includegraphics[angle=0,width=0.8\textwidth]{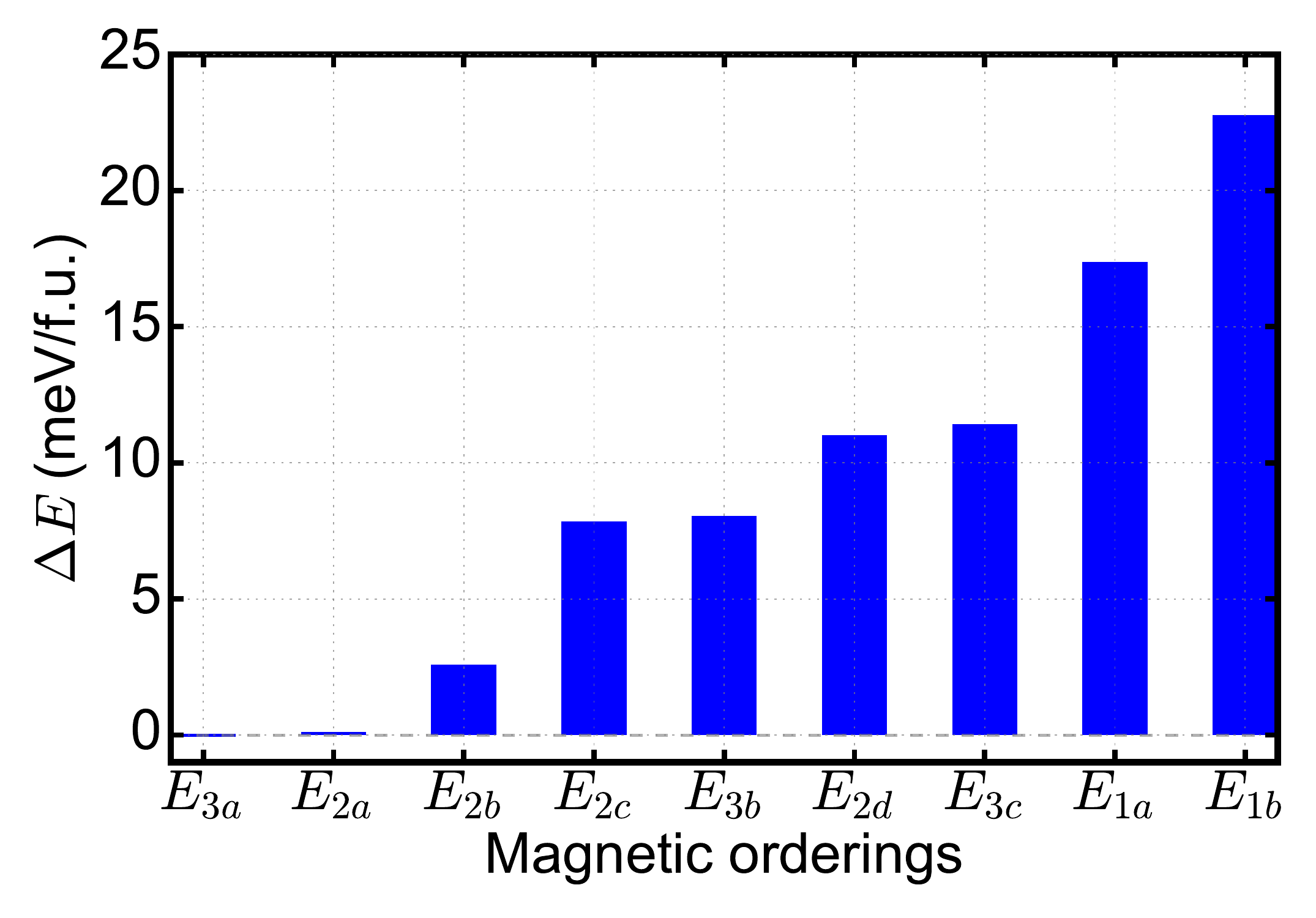}
\caption{\label{fig:fig4} Energy sequence in ascending order of the
  nine stable antiferromagnetic orderings found in Ba$_2$YOsO$_6$ from
  spin-polarized DFT-PBEsol+SOC calculations. The two lowest energy states are a
  non-collinear non-coplanar state $E_{3a}$ and a coplanar state
  $E_{2a}$. The energy of $E_{3a}$ is set as the zero energy.  }
\end{figure}

The effects of SOC are to couple Os $S=3/2$ spins to crystal
lattice. This means that those magnetic orderings which would be
degenerate without SOC now have different energies due to their
different orientations with respect to the crystal lattice, \ie, the
presence of SOC induces anisotropic exchange interaction and leads to
more different complex magnetic
orderings~\cite{YanHan-PRB2017-pyrochlore, Ross2011}. Our
spin-polarized DFT-PBEsol+SOC calculations find that in
Ba$_2$YOsO$_6$, there are nine stable antiferromagnetic orderings,
which are explicitly shown in figure~\ref{fig:fig3}. Similar to the
results of DFT calculations without SOC, we classify these nine
magnetic orderings into three cases: collinear states (2 different
states, labelled as $E_{1a}$ and $E_{1b}$), coplanar states (4
different states, labelled as $E_{2a}$, $E_{2b}$, $E_{2c}$, $E_{2d}$)
and non-collinear non-coplanar states (3 different states, labelled as
$E_{3a}$, $E_{3b}$ and $E_{3c}$). It is straightforward to check that
if we turn off SOC, the two collinear magnetic orderings $E_{1a}$ and
$E_{1b}$ would be degenerate, and the four coplanar magnetic orderings
($E_{2a}$, $E_{2b}$, $E_{2c}$, $E_{2d}$) would also be degenerate in
DFT calculations. Figure~\ref{fig:fig4} shows the energy sequence
sorted in ascending order for these nine stable antiferromagnetic
orderings in Ba$_2$YOsO$_6$. After SOC is taken into account, we find
that the two lowest-energy magnetic orderings are a non-collinear
non-coplanar state $E_{3a}$ and a coplanar state $E_{2a}$, both of
which are more stable than the experimentally observed type-I
collinear ordering ($E_{1a}$ and $E_{1b}$). This result is robust and
does not depended on the choice of exchange-correlation
functionals, the comparison between different exchange-correlation
functionals is provided in section II of Supplementary Materials.

\begin{figure}[t!]
\includegraphics[angle=0,width=0.8\textwidth]{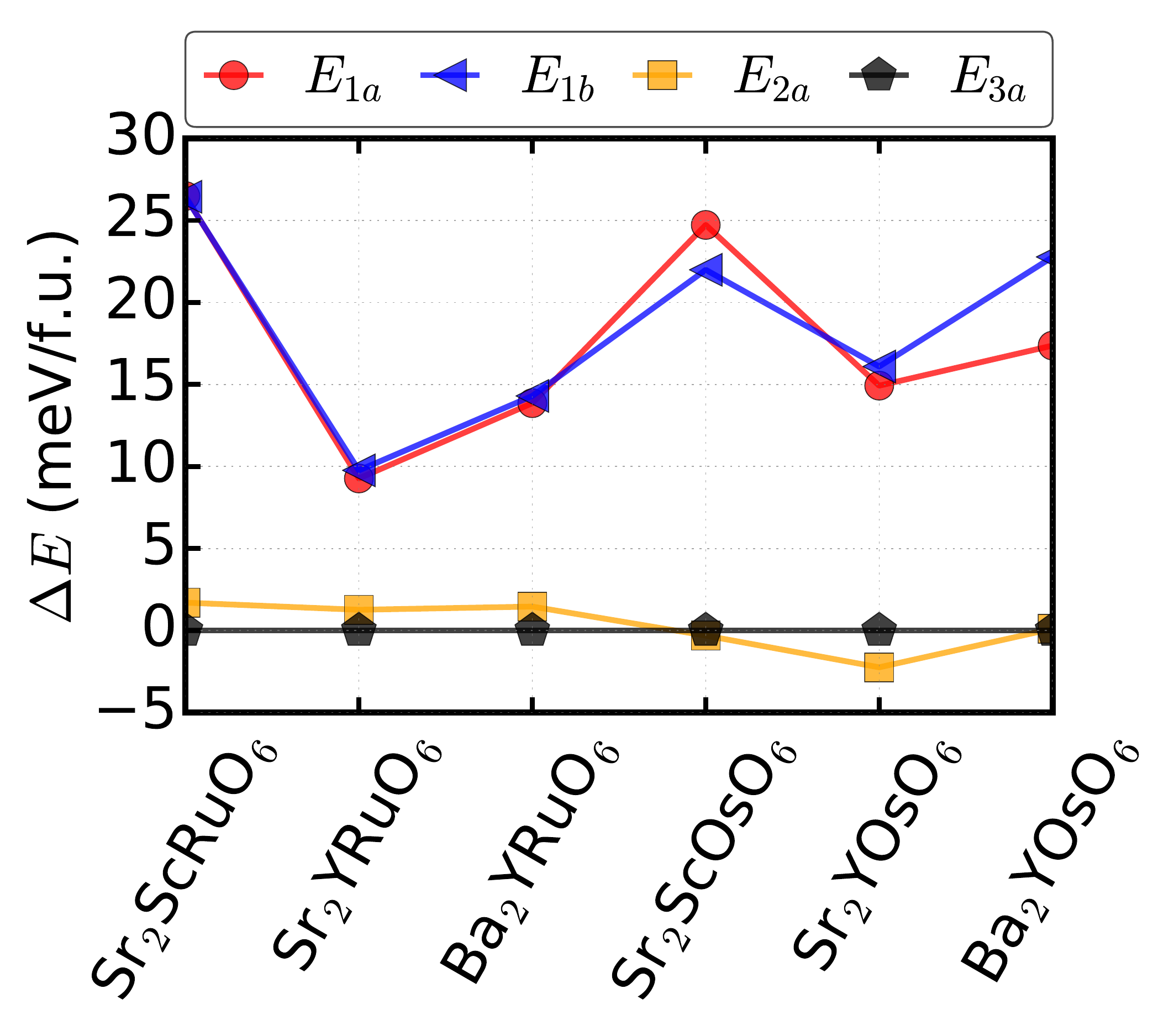}
\caption{\label{fig:verifySOC} The energies of non-collinear
  non-coplanar antiferromagnetic ordering $E_{3a}$, coplanar
  antiferromagnetic ordering $E_{2a}$, and the two
  type-I collinear antiferromagnetic orderings ($E_{1a}$ and $E_{1b}$)
  for all the six complex oxides. Spin-orbit coupling is taken
  into account in the calculations. The energy of the
  noncollinear-noncoplanar $E_{3a}$ is chosen as the zero point.}
\end{figure}

Next, we test the other five
complex oxides and compare the non-collinear non-coplanar
state $E_{3a}$ and the coplanar state $E_{2a}$ to the experimentally
observed type-I collinear ordering ($E_{1a}$ and $E_{1b}$). We
find similar results (see figure~\ref{fig:verifySOC}) that at
zero temperature, the type-I collinear magnetic ordering
is not the most stable one; both the non-collinear non-coplanar
state ($E_{3a}$) and the coplanar state ($E_{2a}$) have lower
energies. This implies that in those complex oxides there could
occur an entropy-driven collinear-to-noncollinear magnetic
transition at sufficiently low temperatures.

Finally, we notice that the ${\Delta E}$ between $E_{2a}$ and $E_{3a}$
of \BYOO~ is extremely small ($\sim$ 0.10 meV/f.u.), indicating these
two antiferromagnetic orderings are almost degenerate.  This is an
accidental degeneracy, which is not protected by symmetry and strongly
depends on materials, as figure~\ref{fig:verifySOC} shows. While
antiferromagnetic orderings are difficult to control due to the
vanishing of net moments, the near-degenerateness between the two
complex antiferromagnetic orderings implies that mechanical strain may
tune Ba$_2$YOsO$_6$ from one magnetic ordering to another.  Here we
consider uniaxial strain, since experimentally amplified piezoelectric
actuators can generate continuously tunable uniaxial strain up to 1\%
~\cite{Hicks-Sr2RuO4-science2014, PRL2018-Sr2RuO4-Barber}. Such
uniaxial strain has been successfully applied to Sr$_2$RuO$_4$ to
enhance its superconducting transition
temperatures~\cite{Steppke-2017Science-Sr2RuO4,Hicks-Sr2RuO4-science2014}
and tune its resistivity in the vicinity of van Hove
singularities~\cite{PRL2018-Sr2RuO4-Barber}. Uniaxial strain has also
been proved to be effective in manipulating the magnetic degrees of freedom
in BaFe$_2$As$_2$~\cite{kissikov2018uniaxial-NC}.

\begin{figure}[t!]
\includegraphics[angle=0,width=0.9\textwidth]{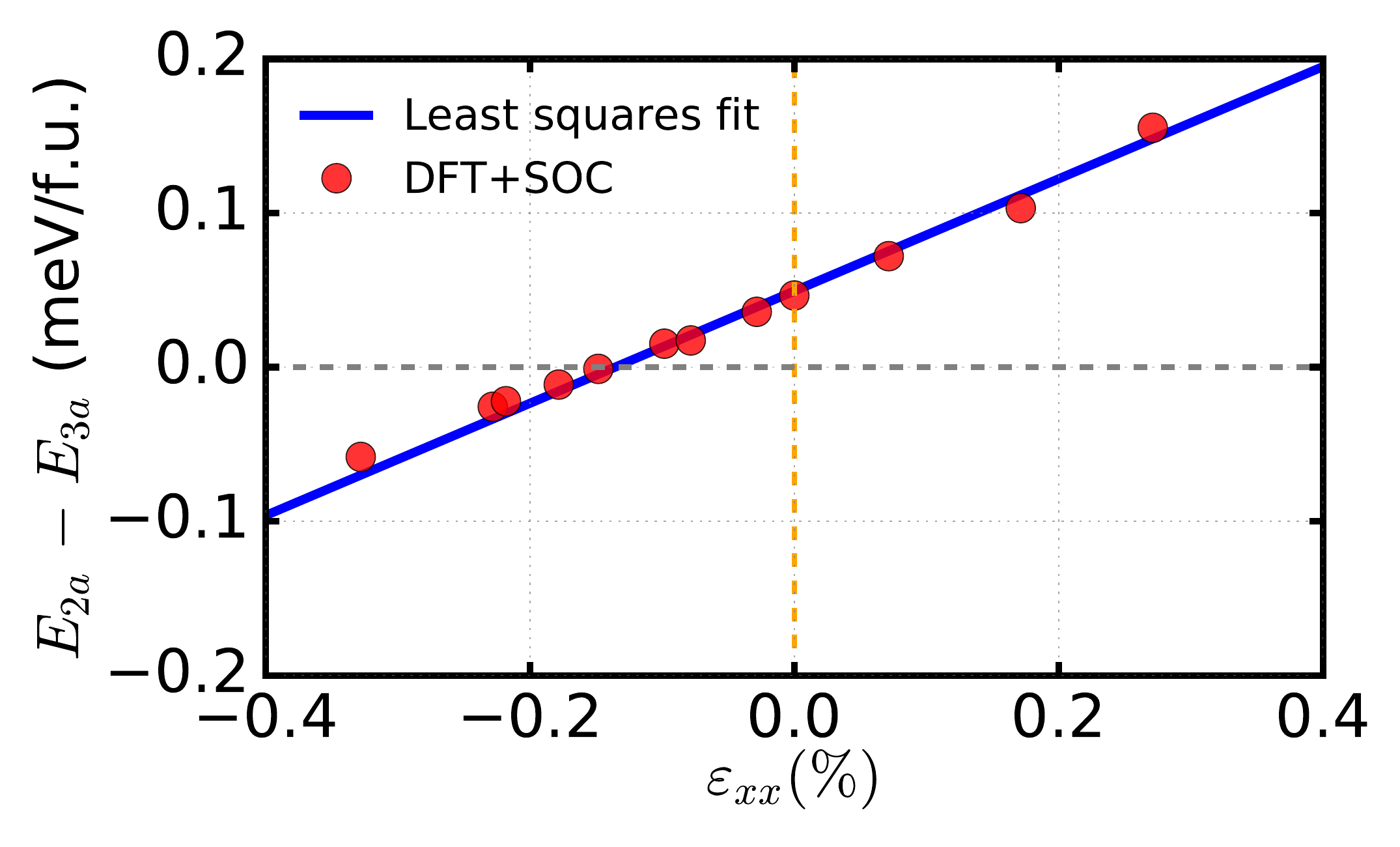}
\caption{\label{fig:socDFT} Energy difference between the coplanar
  state $E_{2a}$ and the non-collinear non-coplanar state $E_{3a}$ as
  a function of uniaxial strain $\varepsilon_{xx}$ (\%).  The red
  circles denote the energy differences obtained from spin-polarized
  DFT-PBEsol+SOC calculations. The blue line is a least squares linear
  fit of the DFT calculated energies.}
\end{figure}

Since we need to study strain effects, we perform structural
optimization for Ba$_2$YOsO$_6$ using spin-polarized DFT-PBEsol+SOC
method and obtain an optimized lattice constant of 8.368~\AA~, which
is in good agreement with the experimental value (within $\sim$ 0.2\%
difference).  We calculate the energy difference between the coplanar
state $E_{2a}$ and the non-collinear non-coplanar state $E_{3a}$ in
Ba$_2$YOsO$_6$ when one of its lattice constant $a_x$ is under strain
while the other two lattice constants as well as internal coordinates
are fully relaxed. Figure~\ref{fig:socDFT} shows $E_{2a} - E_{3a}$ as
a function of uniaxial strain $\varepsilon_{xx}$ which is defined as:
\begin{equation}
\label{eq5} \varepsilon_{xx} = \frac{a_x-a_0}{a_0} \times 100\%
\end{equation}
where $a_0$ is the theoretical lattice constant in equilibrium
obtained from spin-polarized DFT-PBEsol+SOC calculations.  We find
that uniaxial tensile strain favors the non-collinear non-coplanar
state $E_{3a}$; while sufficient uniaxial compressive strain can make
the coplanar state $E_{2a}$ more stable. The critical strain is about
0.1\%~of uniaxial compression, which is achievable by using
piezoelectric
actuators~\cite{Steppke-2017Science-Sr2RuO4,Hicks-Sr2RuO4-science2014,
  PRL2018-Sr2RuO4-Barber}.

\section{Conclusions}

In conclusion, we perform first-principles calculations on a wide
range of magnetically frustrated complex oxides and predict that at
zero temperature, a number of non-collinear magnetic orderings are
more stable than the type-I collinear magnetic ordering that is
observed at finite temperatures. Those non-collinear magnetic
orderings are induced and stabilized by nearest-neighbor biquadratic
interactions, which become pronounced in materials that contain
second-row and third-row transition metal elements.  It implies that
at sufficiently low temperatures, a collinear-to-noncollinear magnetic
transition can occur probably due to entropy effects. We test
different exchange-correlation functionals, various correlation
strengths on magnetic ions and presence/absence of spin-orbit
coupling. We find that our conclusions are robust.

We also demonstrate that a particular complex oxide Ba$_2$YOsO$_6$ is
in the vicinity of the phase boundary between two non-collinear
magnetic orderings. Experimentally feasible uniaxial strain can tune
the material across the magnetic phase boundary, with uniaxial tensile
strain energetically favoring the non-collinear non-coplanar
ordering ($E_{3a}$), and uniaxial compressive strain making the coplanar
ordering ($E_{2a}$) more stable.

Our prediction of zero-temperature 4-sublattice magnetic orderings in
those frustrated complex oxides (with detailed orientation of magnetic
moments) should provide new targets for future neutron scattering
measurements at very low temperatures. Our work also shows that
tunable uniaxial strain can control complex non-collinear magnetic
orderings in Ba$_2$YOsO$_6$, providing an example to studying
spin-lattice interactions in frustrated oxides. We hope our
theoretical calculations can stimulate new experimental study
on magnetic properties of those materials.

\section*{Acknowledgement}
We are grateful to useful discussions with Nic Shannon, Han
Yan, Gang Chen and Wan Yuan. We acknowledge support from National
Natural Science Foundation of China (No. 11774236),
Pujiang Talents program (No. 17PJ1407300), the Seed
Grants of NYU-ECNU Joint Research Institutes and the 2019 University Research Challenge Fund.
Y.-W.F. acknowledges the scholarship granted by the NYU Shanghai
Student Research Program in Physics during his stay at NYU Shanghai.
This research was carried out on the High Performance Computing resources at New York University New York, Abu Dhabi and Shanghai.

\section*{Data availability statement}
The data that support the findings of this study are openly
available in Ref.~\cite{fang-chen-dataset-1}.




\bibliography{DP-SOC}

\begin{thebibliography}{50}
\expandafter\ifx\csname natexlab\endcsname\relax\def\natexlab#1{#1}\fi
\expandafter\ifx\csname bibnamefont\endcsname\relax
  \def\bibnamefont#1{#1}\fi
\expandafter\ifx\csname bibfnamefont\endcsname\relax
  \def\bibfnamefont#1{#1}\fi
\expandafter\ifx\csname citenamefont\endcsname\relax
  \def\citenamefont#1{#1}\fi
\expandafter\ifx\csname url\endcsname\relax
  \def\url#1{\texttt{#1}}\fi
\expandafter\ifx\csname urlprefix\endcsname\relax\def\urlprefix{URL }\fi
\providecommand{\bibinfo}[2]{#2}
\providecommand{\eprint}[2][]{\url{#2}}

\bibitem[{\citenamefont{Balz et~al.}(2016)\citenamefont{Balz, Lake, Reuther,
  Luetkens, Sch{\"o}nemann, Herrmannsd{\"o}rfer, Singh, Islam, Wheeler,
  Rodriguez-Rivera et~al.}}]{balz-natphy-2016VB}
\bibinfo{author}{\bibfnamefont{C.}~\bibnamefont{Balz}},
  \bibinfo{author}{\bibfnamefont{B.}~\bibnamefont{Lake}},
  \bibinfo{author}{\bibfnamefont{J.}~\bibnamefont{Reuther}},
  \bibinfo{author}{\bibfnamefont{H.}~\bibnamefont{Luetkens}},
  \bibinfo{author}{\bibfnamefont{R.}~\bibnamefont{Sch{\"o}nemann}},
  \bibinfo{author}{\bibfnamefont{T.}~\bibnamefont{Herrmannsd{\"o}rfer}},
  \bibinfo{author}{\bibfnamefont{Y.}~\bibnamefont{Singh}},
  \bibinfo{author}{\bibfnamefont{A.~N.} \bibnamefont{Islam}},
  \bibinfo{author}{\bibfnamefont{E.~M.} \bibnamefont{Wheeler}},
  \bibinfo{author}{\bibfnamefont{J.~A.} \bibnamefont{Rodriguez-Rivera}},
  \bibnamefont{et~al.}, \bibinfo{journal}{Nature Phys.}
  \textbf{\bibinfo{volume}{12}}, \bibinfo{pages}{942} (\bibinfo{year}{2016}).

\bibitem[{\citenamefont{Li et~al.}(2017)\citenamefont{Li, Adroja, Voneshen,
  Bewley, Zhang, Tsirlin, and Gegenwart}}]{li-natcomm-2017valencebond}
\bibinfo{author}{\bibfnamefont{Y.}~\bibnamefont{Li}},
  \bibinfo{author}{\bibfnamefont{D.}~\bibnamefont{Adroja}},
  \bibinfo{author}{\bibfnamefont{D.}~\bibnamefont{Voneshen}},
  \bibinfo{author}{\bibfnamefont{R.~I.} \bibnamefont{Bewley}},
  \bibinfo{author}{\bibfnamefont{Q.}~\bibnamefont{Zhang}},
  \bibinfo{author}{\bibfnamefont{A.~A.} \bibnamefont{Tsirlin}},
  \bibnamefont{and}
  \bibinfo{author}{\bibfnamefont{P.}~\bibnamefont{Gegenwart}},
  \bibinfo{journal}{Nat. Commun.} \textbf{\bibinfo{volume}{8}},
  \bibinfo{pages}{15814} (\bibinfo{year}{2017}).

\bibitem[{\citenamefont{Bojesen and Onoda}(2017)}]{Bojesen-PRL-spinice2017}
\bibinfo{author}{\bibfnamefont{T.~A.} \bibnamefont{Bojesen}} \bibnamefont{and}
  \bibinfo{author}{\bibfnamefont{S.}~\bibnamefont{Onoda}},
  \bibinfo{journal}{Phys. Rev. Lett.} \textbf{\bibinfo{volume}{119}},
  \bibinfo{pages}{227204} (\bibinfo{year}{2017}).

\bibitem[{\citenamefont{Lantagne-Hurtubise
  et~al.}(2018)\citenamefont{Lantagne-Hurtubise, Rau, and
  Gingras}}]{PRX-spinice-2018}
\bibinfo{author}{\bibfnamefont{E.}~\bibnamefont{Lantagne-Hurtubise}},
  \bibinfo{author}{\bibfnamefont{J.~G.} \bibnamefont{Rau}}, \bibnamefont{and}
  \bibinfo{author}{\bibfnamefont{M.~J.~P.} \bibnamefont{Gingras}},
  \bibinfo{journal}{Phys. Rev. X} \textbf{\bibinfo{volume}{8}},
  \bibinfo{pages}{021053} (\bibinfo{year}{2018}).

\bibitem[{\citenamefont{Balents}(2010)}]{balents-nature-2010spinliquid}
\bibinfo{author}{\bibfnamefont{L.}~\bibnamefont{Balents}},
  \bibinfo{journal}{Nature} \textbf{\bibinfo{volume}{464}},
  \bibinfo{pages}{199} (\bibinfo{year}{2010}).

\bibitem[{\citenamefont{Rousochatzakis
  et~al.}(2018)\citenamefont{Rousochatzakis, Sizyuk, and
  Perkins}}]{rousochatzakis-natcomm-spinliquid2018}
\bibinfo{author}{\bibfnamefont{I.}~\bibnamefont{Rousochatzakis}},
  \bibinfo{author}{\bibfnamefont{Y.}~\bibnamefont{Sizyuk}}, \bibnamefont{and}
  \bibinfo{author}{\bibfnamefont{N.~B.} \bibnamefont{Perkins}},
  \bibinfo{journal}{Nat. Commun.} \textbf{\bibinfo{volume}{9}},
  \bibinfo{pages}{1575} (\bibinfo{year}{2018}).

\bibitem[{\citenamefont{Savary and Balents}(2016)}]{savary2016Leon-review}
\bibinfo{author}{\bibfnamefont{L.}~\bibnamefont{Savary}} \bibnamefont{and}
  \bibinfo{author}{\bibfnamefont{L.}~\bibnamefont{Balents}},
  \bibinfo{journal}{Rep. Prog. Phys.} \textbf{\bibinfo{volume}{80}},
  \bibinfo{pages}{016502} (\bibinfo{year}{2016}).

\bibitem[{\citenamefont{Zhang et~al.}(2017)\citenamefont{Zhang, Baker, Zhang,
  Wang, Wang, Su, Zhu, and Pratt}}]{zhang2017quantum}
\bibinfo{author}{\bibfnamefont{B.}~\bibnamefont{Zhang}},
  \bibinfo{author}{\bibfnamefont{P.~J.} \bibnamefont{Baker}},
  \bibinfo{author}{\bibfnamefont{Y.}~\bibnamefont{Zhang}},
  \bibinfo{author}{\bibfnamefont{D.}~\bibnamefont{Wang}},
  \bibinfo{author}{\bibfnamefont{Z.}~\bibnamefont{Wang}},
  \bibinfo{author}{\bibfnamefont{S.}~\bibnamefont{Su}},
  \bibinfo{author}{\bibfnamefont{D.}~\bibnamefont{Zhu}}, \bibnamefont{and}
  \bibinfo{author}{\bibfnamefont{F.~L.} \bibnamefont{Pratt}},
  \bibinfo{journal}{J. Am. Chem. Soc.} \textbf{\bibinfo{volume}{140}},
  \bibinfo{pages}{122} (\bibinfo{year}{2017}).

\bibitem[{\citenamefont{Norman}(2016)}]{RevModPhys.88.041002}
\bibinfo{author}{\bibfnamefont{M.~R.} \bibnamefont{Norman}},
  \bibinfo{journal}{Rev. Mod. Phys.} \textbf{\bibinfo{volume}{88}},
  \bibinfo{pages}{041002} (\bibinfo{year}{2016}).

\bibitem[{\citenamefont{Kayser et~al.}(2017{\natexlab{a}})\citenamefont{Kayser,
  Kennedy, Ranjbar, Kimpton, and Avdeev}}]{kayser2017spin-organicchem}
\bibinfo{author}{\bibfnamefont{P.}~\bibnamefont{Kayser}},
  \bibinfo{author}{\bibfnamefont{B.~J.} \bibnamefont{Kennedy}},
  \bibinfo{author}{\bibfnamefont{B.}~\bibnamefont{Ranjbar}},
  \bibinfo{author}{\bibfnamefont{J.~A.} \bibnamefont{Kimpton}},
  \bibnamefont{and} \bibinfo{author}{\bibfnamefont{M.}~\bibnamefont{Avdeev}},
  \bibinfo{journal}{Inorg. Chem.} \textbf{\bibinfo{volume}{56}},
  \bibinfo{pages}{2204} (\bibinfo{year}{2017}{\natexlab{a}}).

\bibitem[{\citenamefont{Taylor et~al.}(2015)\citenamefont{Taylor, Morrow,
  Singh, Calder, Lumsden, Woodward, and Christianson}}]{PhysRevB.91.100406}
\bibinfo{author}{\bibfnamefont{A.~E.} \bibnamefont{Taylor}},
  \bibinfo{author}{\bibfnamefont{R.}~\bibnamefont{Morrow}},
  \bibinfo{author}{\bibfnamefont{D.~J.} \bibnamefont{Singh}},
  \bibinfo{author}{\bibfnamefont{S.}~\bibnamefont{Calder}},
  \bibinfo{author}{\bibfnamefont{M.~D.} \bibnamefont{Lumsden}},
  \bibinfo{author}{\bibfnamefont{P.~M.} \bibnamefont{Woodward}},
  \bibnamefont{and} \bibinfo{author}{\bibfnamefont{A.~D.}
  \bibnamefont{Christianson}}, \bibinfo{journal}{Phys. Rev. B}
  \textbf{\bibinfo{volume}{91}}, \bibinfo{pages}{100406}
  (\bibinfo{year}{2015}).

\bibitem[{\citenamefont{Taylor et~al.}(2016)\citenamefont{Taylor, Morrow,
  Fishman, Calder, Kolesnikov, Lumsden, Woodward, and
  Christianson}}]{PhysRevB.93.220408}
\bibinfo{author}{\bibfnamefont{A.~E.} \bibnamefont{Taylor}},
  \bibinfo{author}{\bibfnamefont{R.}~\bibnamefont{Morrow}},
  \bibinfo{author}{\bibfnamefont{R.~S.} \bibnamefont{Fishman}},
  \bibinfo{author}{\bibfnamefont{S.}~\bibnamefont{Calder}},
  \bibinfo{author}{\bibfnamefont{A.~I.} \bibnamefont{Kolesnikov}},
  \bibinfo{author}{\bibfnamefont{M.~D.} \bibnamefont{Lumsden}},
  \bibinfo{author}{\bibfnamefont{P.~M.} \bibnamefont{Woodward}},
  \bibnamefont{and} \bibinfo{author}{\bibfnamefont{A.~D.}
  \bibnamefont{Christianson}}, \bibinfo{journal}{Phys. Rev. B}
  \textbf{\bibinfo{volume}{93}}, \bibinfo{pages}{220408}
  (\bibinfo{year}{2016}).

\bibitem[{\citenamefont{Granado et~al.}(2013)\citenamefont{Granado, Lynn,
  Jardim, and Torikachvili}}]{PhysRevLett.110.017202}
\bibinfo{author}{\bibfnamefont{E.}~\bibnamefont{Granado}},
  \bibinfo{author}{\bibfnamefont{J.~W.} \bibnamefont{Lynn}},
  \bibinfo{author}{\bibfnamefont{R.~F.} \bibnamefont{Jardim}},
  \bibnamefont{and} \bibinfo{author}{\bibfnamefont{M.~S.}
  \bibnamefont{Torikachvili}}, \bibinfo{journal}{Phys. Rev. Lett.}
  \textbf{\bibinfo{volume}{110}}, \bibinfo{pages}{017202}
  (\bibinfo{year}{2013}).

\bibitem[{\citenamefont{Nilsen et~al.}(2015)\citenamefont{Nilsen, Thompson,
  Ehlers, Marjerrison, and Greedan}}]{PhysRevB.91.054415}
\bibinfo{author}{\bibfnamefont{G.~J.} \bibnamefont{Nilsen}},
  \bibinfo{author}{\bibfnamefont{C.~M.} \bibnamefont{Thompson}},
  \bibinfo{author}{\bibfnamefont{G.}~\bibnamefont{Ehlers}},
  \bibinfo{author}{\bibfnamefont{C.~A.} \bibnamefont{Marjerrison}},
  \bibnamefont{and} \bibinfo{author}{\bibfnamefont{J.~E.}
  \bibnamefont{Greedan}}, \bibinfo{journal}{Phys. Rev. B}
  \textbf{\bibinfo{volume}{91}}, \bibinfo{pages}{054415}
  (\bibinfo{year}{2015}).

\bibitem[{\citenamefont{Paul et~al.}(2015)\citenamefont{Paul, Sarapulova,
  Adler, Reehuis, Kanungo, Mikhailova, Schnelle, Hu, Kuo, Siruguri
  et~al.}}]{ZAAC:ZAAC201400590}
\bibinfo{author}{\bibfnamefont{A.~K.} \bibnamefont{Paul}},
  \bibinfo{author}{\bibfnamefont{A.}~\bibnamefont{Sarapulova}},
  \bibinfo{author}{\bibfnamefont{P.}~\bibnamefont{Adler}},
  \bibinfo{author}{\bibfnamefont{M.}~\bibnamefont{Reehuis}},
  \bibinfo{author}{\bibfnamefont{S.}~\bibnamefont{Kanungo}},
  \bibinfo{author}{\bibfnamefont{D.}~\bibnamefont{Mikhailova}},
  \bibinfo{author}{\bibfnamefont{W.}~\bibnamefont{Schnelle}},
  \bibinfo{author}{\bibfnamefont{Z.}~\bibnamefont{Hu}},
  \bibinfo{author}{\bibfnamefont{C.}~\bibnamefont{Kuo}},
  \bibinfo{author}{\bibfnamefont{V.}~\bibnamefont{Siruguri}},
  \bibnamefont{et~al.}, \bibinfo{journal}{Zeitschrift für anorganische und
  allgemeine Chemie} \textbf{\bibinfo{volume}{641}}, \bibinfo{pages}{197}
  (\bibinfo{year}{2015}).

\bibitem[{\citenamefont{Maharaj et~al.}(2018)\citenamefont{Maharaj, Sala,
  Marjerrison, Stone, Greedan, and Gaulin}}]{PhysRevB.98.104434}
\bibinfo{author}{\bibfnamefont{D.~D.} \bibnamefont{Maharaj}},
  \bibinfo{author}{\bibfnamefont{G.}~\bibnamefont{Sala}},
  \bibinfo{author}{\bibfnamefont{C.~A.} \bibnamefont{Marjerrison}},
  \bibinfo{author}{\bibfnamefont{M.~B.} \bibnamefont{Stone}},
  \bibinfo{author}{\bibfnamefont{J.~E.} \bibnamefont{Greedan}},
  \bibnamefont{and} \bibinfo{author}{\bibfnamefont{B.~D.}
  \bibnamefont{Gaulin}}, \bibinfo{journal}{Phys. Rev. B}
  \textbf{\bibinfo{volume}{98}}, \bibinfo{pages}{104434}
  (\bibinfo{year}{2018}).

\bibitem[{\citenamefont{Kayser et~al.}(2017{\natexlab{b}})\citenamefont{Kayser,
  Injac, Ranjbar, Kennedy, Avdeev, and
  Yamaura}}]{doi:10.1021/acs.inorgchem.7b00983}
\bibinfo{author}{\bibfnamefont{P.}~\bibnamefont{Kayser}},
  \bibinfo{author}{\bibfnamefont{S.}~\bibnamefont{Injac}},
  \bibinfo{author}{\bibfnamefont{B.}~\bibnamefont{Ranjbar}},
  \bibinfo{author}{\bibfnamefont{B.~J.} \bibnamefont{Kennedy}},
  \bibinfo{author}{\bibfnamefont{M.}~\bibnamefont{Avdeev}}, \bibnamefont{and}
  \bibinfo{author}{\bibfnamefont{K.}~\bibnamefont{Yamaura}},
  \bibinfo{journal}{Inorg. Chem.} \textbf{\bibinfo{volume}{56}},
  \bibinfo{pages}{9009} (\bibinfo{year}{2017}{\natexlab{b}}).

\bibitem[{\citenamefont{Battle and Macklin}(1984)}]{BATTLE1984138}
\bibinfo{author}{\bibfnamefont{P.}~\bibnamefont{Battle}} \bibnamefont{and}
  \bibinfo{author}{\bibfnamefont{W.}~\bibnamefont{Macklin}},
  \bibinfo{journal}{J. Solid State Chem.} \textbf{\bibinfo{volume}{52}},
  \bibinfo{pages}{138 } (\bibinfo{year}{1984}).

\bibitem[{\citenamefont{Izumiyama et~al.}(2002)\citenamefont{Izumiyama, Doi,
  Wakeshima, Hinatsu, Nakamura, and Ishii}}]{IZUMIYAMA2002125}
\bibinfo{author}{\bibfnamefont{Y.}~\bibnamefont{Izumiyama}},
  \bibinfo{author}{\bibfnamefont{Y.}~\bibnamefont{Doi}},
  \bibinfo{author}{\bibfnamefont{M.}~\bibnamefont{Wakeshima}},
  \bibinfo{author}{\bibfnamefont{Y.}~\bibnamefont{Hinatsu}},
  \bibinfo{author}{\bibfnamefont{A.}~\bibnamefont{Nakamura}}, \bibnamefont{and}
  \bibinfo{author}{\bibfnamefont{Y.}~\bibnamefont{Ishii}}, \bibinfo{journal}{J.
  Solid State Chem.} \textbf{\bibinfo{volume}{169}}, \bibinfo{pages}{125 }
  (\bibinfo{year}{2002}).

\bibitem[{\citenamefont{Kermarrec et~al.}(2015)\citenamefont{Kermarrec,
  Marjerrison, Thompson, Maharaj, Levin, Kroeker, Granroth, Flacau, Yamani,
  Greedan et~al.}}]{Kermarrec-BYOO-PRB2015}
\bibinfo{author}{\bibfnamefont{E.}~\bibnamefont{Kermarrec}},
  \bibinfo{author}{\bibfnamefont{C.~A.} \bibnamefont{Marjerrison}},
  \bibinfo{author}{\bibfnamefont{C.~M.} \bibnamefont{Thompson}},
  \bibinfo{author}{\bibfnamefont{D.~D.} \bibnamefont{Maharaj}},
  \bibinfo{author}{\bibfnamefont{K.}~\bibnamefont{Levin}},
  \bibinfo{author}{\bibfnamefont{S.}~\bibnamefont{Kroeker}},
  \bibinfo{author}{\bibfnamefont{G.~E.} \bibnamefont{Granroth}},
  \bibinfo{author}{\bibfnamefont{R.}~\bibnamefont{Flacau}},
  \bibinfo{author}{\bibfnamefont{Z.}~\bibnamefont{Yamani}},
  \bibinfo{author}{\bibfnamefont{J.~E.} \bibnamefont{Greedan}},
  \bibnamefont{et~al.}, \bibinfo{journal}{Phys. Rev. B}
  \textbf{\bibinfo{volume}{91}}, \bibinfo{pages}{075133}
  (\bibinfo{year}{2015}).

\bibitem[{\citenamefont{Battle and Jones}(1989)}]{BATTLE1989108}
\bibinfo{author}{\bibfnamefont{P.}~\bibnamefont{Battle}} \bibnamefont{and}
  \bibinfo{author}{\bibfnamefont{C.}~\bibnamefont{Jones}}, \bibinfo{journal}{J.
  Solid State Chem.} \textbf{\bibinfo{volume}{78}}, \bibinfo{pages}{108 }
  (\bibinfo{year}{1989}).

\bibitem[{\citenamefont{Kresse and Furthm$\ddot{\rm
  u}$ller}(1996{\natexlab{a}})}]{Kresse1996}
\bibinfo{author}{\bibfnamefont{G.}~\bibnamefont{Kresse}} \bibnamefont{and}
  \bibinfo{author}{\bibfnamefont{J.}~\bibnamefont{Furthm$\ddot{\rm u}$ller}},
  \bibinfo{journal}{Comp. Mater. Sci.} \textbf{\bibinfo{volume}{6}},
  \bibinfo{pages}{15 } (\bibinfo{year}{1996}{\natexlab{a}}).

\bibitem[{\citenamefont{Kresse and Furthm$\ddot{\rm
  u}$ller}(1996{\natexlab{b}})}]{Kresse-PRB-1996}
\bibinfo{author}{\bibfnamefont{G.}~\bibnamefont{Kresse}} \bibnamefont{and}
  \bibinfo{author}{\bibfnamefont{J.}~\bibnamefont{Furthm$\ddot{\rm u}$ller}},
  \bibinfo{journal}{Phys. Rev. B} \textbf{\bibinfo{volume}{54}},
  \bibinfo{pages}{11169} (\bibinfo{year}{1996}{\natexlab{b}}).

\bibitem[{\citenamefont{Dudarev et~al.}(1998)\citenamefont{Dudarev, Botton,
  Savrasov, Humphreys, and Sutton}}]{Dudarev-LDAU-PRB1998}
\bibinfo{author}{\bibfnamefont{S.~L.} \bibnamefont{Dudarev}},
  \bibinfo{author}{\bibfnamefont{G.~A.} \bibnamefont{Botton}},
  \bibinfo{author}{\bibfnamefont{S.~Y.} \bibnamefont{Savrasov}},
  \bibinfo{author}{\bibfnamefont{C.~J.} \bibnamefont{Humphreys}},
  \bibnamefont{and} \bibinfo{author}{\bibfnamefont{A.~P.}
  \bibnamefont{Sutton}}, \bibinfo{journal}{Phys. Rev. B}
  \textbf{\bibinfo{volume}{57}}, \bibinfo{pages}{1505} (\bibinfo{year}{1998}).

\bibitem[{\citenamefont{Perdew et~al.}(2008)\citenamefont{Perdew, Ruzsinszky,
  Csonka, Vydrov, Scuseria, Constantin, Zhou, and
  Burke}}]{PhysRevLett.100.136406PBEsol}
\bibinfo{author}{\bibfnamefont{J.~P.} \bibnamefont{Perdew}},
  \bibinfo{author}{\bibfnamefont{A.}~\bibnamefont{Ruzsinszky}},
  \bibinfo{author}{\bibfnamefont{G.~I.} \bibnamefont{Csonka}},
  \bibinfo{author}{\bibfnamefont{O.~A.} \bibnamefont{Vydrov}},
  \bibinfo{author}{\bibfnamefont{G.~E.} \bibnamefont{Scuseria}},
  \bibinfo{author}{\bibfnamefont{L.~A.} \bibnamefont{Constantin}},
  \bibinfo{author}{\bibfnamefont{X.}~\bibnamefont{Zhou}}, \bibnamefont{and}
  \bibinfo{author}{\bibfnamefont{K.}~\bibnamefont{Burke}},
  \bibinfo{journal}{Phys. Rev. Lett.} \textbf{\bibinfo{volume}{100}},
  \bibinfo{pages}{136406} (\bibinfo{year}{2008}).

\bibitem[{\citenamefont{Chen}(2018)}]{Chen_npj2018}
\bibinfo{author}{\bibfnamefont{H.}~\bibnamefont{Chen}}, \bibinfo{journal}{npj
  Quant. Mater.} \textbf{\bibinfo{volume}{3}}, \bibinfo{pages}{57}
  (\bibinfo{year}{2018}).

\bibitem[{\citenamefont{Aulesti et~al.}(2018)\citenamefont{Aulesti, Cheung,
  Fang, He, Yamaura, Lai, Goh, and Chen}}]{aulesti2018APL}
\bibinfo{author}{\bibfnamefont{E.~I.~P.} \bibnamefont{Aulesti}},
  \bibinfo{author}{\bibfnamefont{Y.~W.} \bibnamefont{Cheung}},
  \bibinfo{author}{\bibfnamefont{Y.-W.} \bibnamefont{Fang}},
  \bibinfo{author}{\bibfnamefont{J.}~\bibnamefont{He}},
  \bibinfo{author}{\bibfnamefont{K.}~\bibnamefont{Yamaura}},
  \bibinfo{author}{\bibfnamefont{K.~T.} \bibnamefont{Lai}},
  \bibinfo{author}{\bibfnamefont{S.~K.} \bibnamefont{Goh}}, \bibnamefont{and}
  \bibinfo{author}{\bibfnamefont{H.}~\bibnamefont{Chen}},
  \bibinfo{journal}{Appl. Phys. Lett.} \textbf{\bibinfo{volume}{113}},
  \bibinfo{pages}{12902} (\bibinfo{year}{2018}).

\bibitem[{\citenamefont{Treiber and
  Kemmler-Sack}(1981)}]{BYOO-expstructure-1981-german}
\bibinfo{author}{\bibfnamefont{U.}~\bibnamefont{Treiber}} \bibnamefont{and}
  \bibinfo{author}{\bibfnamefont{S.}~\bibnamefont{Kemmler-Sack}},
  \bibinfo{journal}{Z. anorg. allg. Chem. (ZAAC)}
  \textbf{\bibinfo{volume}{478}}, \bibinfo{pages}{223} (\bibinfo{year}{1981}).

\bibitem[{\citenamefont{Chen and Balents}(2011)}]{ChenGang_PRB2011dp-d2}
\bibinfo{author}{\bibfnamefont{G.}~\bibnamefont{Chen}} \bibnamefont{and}
  \bibinfo{author}{\bibfnamefont{L.}~\bibnamefont{Balents}},
  \bibinfo{journal}{Phys. Rev. B} \textbf{\bibinfo{volume}{84}},
  \bibinfo{pages}{094420} (\bibinfo{year}{2011}).

\bibitem[{\citenamefont{Lan et~al.}(2018)\citenamefont{Lan, Song, and
  Yang}}]{LAN2018909}
\bibinfo{author}{\bibfnamefont{G.}~\bibnamefont{Lan}},
  \bibinfo{author}{\bibfnamefont{J.}~\bibnamefont{Song}}, \bibnamefont{and}
  \bibinfo{author}{\bibfnamefont{Z.}~\bibnamefont{Yang}}, \bibinfo{journal}{J.
  Alloy. Compd.} \textbf{\bibinfo{volume}{749}}, \bibinfo{pages}{909 }
  (\bibinfo{year}{2018}).

\bibitem[{\citenamefont{Kanungo et~al.}(2014)\citenamefont{Kanungo, Yan,
  Jansen, and Felser}}]{PhysRevB.89.214414}
\bibinfo{author}{\bibfnamefont{S.}~\bibnamefont{Kanungo}},
  \bibinfo{author}{\bibfnamefont{B.}~\bibnamefont{Yan}},
  \bibinfo{author}{\bibfnamefont{M.}~\bibnamefont{Jansen}}, \bibnamefont{and}
  \bibinfo{author}{\bibfnamefont{C.}~\bibnamefont{Felser}},
  \bibinfo{journal}{Phys. Rev. B} \textbf{\bibinfo{volume}{89}},
  \bibinfo{pages}{214414} (\bibinfo{year}{2014}).

\bibitem[{\citenamefont{Chen et~al.}(2015)\citenamefont{Chen, Millis, and
  Marianetti}}]{JiaChen2015}
\bibinfo{author}{\bibfnamefont{J.}~\bibnamefont{Chen}},
  \bibinfo{author}{\bibfnamefont{A.~J.} \bibnamefont{Millis}},
  \bibnamefont{and} \bibinfo{author}{\bibfnamefont{C.~A.}
  \bibnamefont{Marianetti}}, \bibinfo{journal}{Phys. Rev. B}
  \textbf{\bibinfo{volume}{91}}, \bibinfo{pages}{241111}
  (\bibinfo{year}{2015}).

\bibitem[{\citenamefont{Chen and Millis}(2016)}]{HChen-2016-PRB}
\bibinfo{author}{\bibfnamefont{H.}~\bibnamefont{Chen}} \bibnamefont{and}
  \bibinfo{author}{\bibfnamefont{A.~J.} \bibnamefont{Millis}},
  \bibinfo{journal}{Phys. Rev. B} \textbf{\bibinfo{volume}{93}},
  \bibinfo{pages}{205110} (\bibinfo{year}{2016}).

\bibitem[{\citenamefont{Oguchi et~al.}(1985)\citenamefont{Oguchi, Nishimori,
  and Taguchi}}]{Oguchi1985}
\bibinfo{author}{\bibfnamefont{T.}~\bibnamefont{Oguchi}},
  \bibinfo{author}{\bibfnamefont{H.}~\bibnamefont{Nishimori}},
  \bibnamefont{and} \bibinfo{author}{\bibfnamefont{Y.}~\bibnamefont{Taguchi}},
  \bibinfo{journal}{J. Phys. Soc. Jpn.} \textbf{\bibinfo{volume}{54}},
  \bibinfo{pages}{4494} (\bibinfo{year}{1985}).

\bibitem[{\citenamefont{Calzado et~al.}(2003)\citenamefont{Calzado, de~Graaf,
  Bordas, Caballol, and Malrieu}}]{Calzado2003}
\bibinfo{author}{\bibfnamefont{C.~J.} \bibnamefont{Calzado}},
  \bibinfo{author}{\bibfnamefont{C.}~\bibnamefont{de~Graaf}},
  \bibinfo{author}{\bibfnamefont{E.}~\bibnamefont{Bordas}},
  \bibinfo{author}{\bibfnamefont{R.}~\bibnamefont{Caballol}}, \bibnamefont{and}
  \bibinfo{author}{\bibfnamefont{J.-P.} \bibnamefont{Malrieu}},
  \bibinfo{journal}{Phys. Rev. B} \textbf{\bibinfo{volume}{67}},
  \bibinfo{pages}{132409} (\bibinfo{year}{2003}).

\bibitem[{\citenamefont{Coldea et~al.}(2001)\citenamefont{Coldea, Hayden,
  Aeppli, Perring, Frost, Mason, Cheong, and Fisk}}]{Coldea2001}
\bibinfo{author}{\bibfnamefont{R.}~\bibnamefont{Coldea}},
  \bibinfo{author}{\bibfnamefont{S.~M.} \bibnamefont{Hayden}},
  \bibinfo{author}{\bibfnamefont{G.}~\bibnamefont{Aeppli}},
  \bibinfo{author}{\bibfnamefont{T.~G.} \bibnamefont{Perring}},
  \bibinfo{author}{\bibfnamefont{C.~D.} \bibnamefont{Frost}},
  \bibinfo{author}{\bibfnamefont{T.~E.} \bibnamefont{Mason}},
  \bibinfo{author}{\bibfnamefont{S.-W.} \bibnamefont{Cheong}},
  \bibnamefont{and} \bibinfo{author}{\bibfnamefont{Z.}~\bibnamefont{Fisk}},
  \bibinfo{journal}{Phys. Rev. Lett.} \textbf{\bibinfo{volume}{86}},
  \bibinfo{pages}{5377} (\bibinfo{year}{2001}).

\bibitem[{\citenamefont{Xiang et~al.}(2013)\citenamefont{Xiang, Lee, Koo, Gong,
  and Whangbo}}]{Xiang2013}
\bibinfo{author}{\bibfnamefont{H.}~\bibnamefont{Xiang}},
  \bibinfo{author}{\bibfnamefont{C.}~\bibnamefont{Lee}},
  \bibinfo{author}{\bibfnamefont{H.-J.} \bibnamefont{Koo}},
  \bibinfo{author}{\bibfnamefont{X.}~\bibnamefont{Gong}}, \bibnamefont{and}
  \bibinfo{author}{\bibfnamefont{M.-H.} \bibnamefont{Whangbo}},
  \bibinfo{journal}{Dalton Trans.} \textbf{\bibinfo{volume}{42}},
  \bibinfo{pages}{823} (\bibinfo{year}{2013}).

\bibitem[{\citenamefont{Fedorova et~al.}(2015)\citenamefont{Fedorova, Ederer,
  Spaldin, and Scaramucci}}]{Fedorova2015}
\bibinfo{author}{\bibfnamefont{N.~S.} \bibnamefont{Fedorova}},
  \bibinfo{author}{\bibfnamefont{C.}~\bibnamefont{Ederer}},
  \bibinfo{author}{\bibfnamefont{N.~A.} \bibnamefont{Spaldin}},
  \bibnamefont{and}
  \bibinfo{author}{\bibfnamefont{A.}~\bibnamefont{Scaramucci}},
  \bibinfo{journal}{Phys. Rev. B} \textbf{\bibinfo{volume}{91}},
  \bibinfo{pages}{165122} (\bibinfo{year}{2015}).

\bibitem[{\citenamefont{Penc et~al.}(2004)\citenamefont{Penc, Shannon, and
  Shiba}}]{Penc2004}
\bibinfo{author}{\bibfnamefont{K.}~\bibnamefont{Penc}},
  \bibinfo{author}{\bibfnamefont{N.}~\bibnamefont{Shannon}}, \bibnamefont{and}
  \bibinfo{author}{\bibfnamefont{H.}~\bibnamefont{Shiba}},
  \bibinfo{journal}{Phys. Rev. Lett.} \textbf{\bibinfo{volume}{93}},
  \bibinfo{pages}{197203} (\bibinfo{year}{2004}).

\bibitem[{\citenamefont{Penc et~al.}(2007)\citenamefont{Penc, Shannon, Motome,
  and Shiba}}]{Penc2007}
\bibinfo{author}{\bibfnamefont{K.}~\bibnamefont{Penc}},
  \bibinfo{author}{\bibfnamefont{N.}~\bibnamefont{Shannon}},
  \bibinfo{author}{\bibfnamefont{Y.}~\bibnamefont{Motome}}, \bibnamefont{and}
  \bibinfo{author}{\bibfnamefont{H.}~\bibnamefont{Shiba}}, \bibinfo{journal}{J.
  Phys. Condens. Matter} \textbf{\bibinfo{volume}{19}}, \bibinfo{pages}{145267}
  (\bibinfo{year}{2007}).

\bibitem[{\citenamefont{Shannon et~al.}(2010)\citenamefont{Shannon, Penc, and
  Motome}}]{Shannon2010}
\bibinfo{author}{\bibfnamefont{N.}~\bibnamefont{Shannon}},
  \bibinfo{author}{\bibfnamefont{K.}~\bibnamefont{Penc}}, \bibnamefont{and}
  \bibinfo{author}{\bibfnamefont{Y.}~\bibnamefont{Motome}},
  \bibinfo{journal}{Phys. Rev. B} \textbf{\bibinfo{volume}{81}},
  \bibinfo{pages}{184409} (\bibinfo{year}{2010}).

\bibitem[{\citenamefont{Kaplan}(2009)}]{Kaplan2009}
\bibinfo{author}{\bibfnamefont{T.~A.} \bibnamefont{Kaplan}},
  \bibinfo{journal}{Phys. Rev. B} \textbf{\bibinfo{volume}{80}},
  \bibinfo{pages}{012407} (\bibinfo{year}{2009}).

\bibitem[{\citenamefont{Takahashi}(1977)}]{Takahashi1977}
\bibinfo{author}{\bibfnamefont{M.}~\bibnamefont{Takahashi}},
  \bibinfo{journal}{J. Phys. C: Solid State Phys.}
  \textbf{\bibinfo{volume}{10}}, \bibinfo{pages}{1289} (\bibinfo{year}{1977}).

\bibitem[{\citenamefont{Yan et~al.}(2017)\citenamefont{Yan, Benton, Jaubert,
  and Shannon}}]{YanHan-PRB2017-pyrochlore}
\bibinfo{author}{\bibfnamefont{H.}~\bibnamefont{Yan}},
  \bibinfo{author}{\bibfnamefont{O.}~\bibnamefont{Benton}},
  \bibinfo{author}{\bibfnamefont{L.}~\bibnamefont{Jaubert}}, \bibnamefont{and}
  \bibinfo{author}{\bibfnamefont{N.}~\bibnamefont{Shannon}},
  \bibinfo{journal}{Phys. Rev. B} \textbf{\bibinfo{volume}{95}},
  \bibinfo{pages}{094422} (\bibinfo{year}{2017}).

\bibitem[{\citenamefont{Ross et~al.}(2011)\citenamefont{Ross, Savary, Gaulin,
  and Balents}}]{Ross2011}
\bibinfo{author}{\bibfnamefont{K.~A.} \bibnamefont{Ross}},
  \bibinfo{author}{\bibfnamefont{L.}~\bibnamefont{Savary}},
  \bibinfo{author}{\bibfnamefont{B.~D.} \bibnamefont{Gaulin}},
  \bibnamefont{and} \bibinfo{author}{\bibfnamefont{L.}~\bibnamefont{Balents}},
  \bibinfo{journal}{Phys. Rev. X} \textbf{\bibinfo{volume}{1}},
  \bibinfo{pages}{021002} (\bibinfo{year}{2011}).

\bibitem[{\citenamefont{Hicks et~al.}(2014)\citenamefont{Hicks, Brodsky,
  Yelland, Gibbs, Bruin, Barber, Edkins, Nishimura, Yonezawa, Maeno
  et~al.}}]{Hicks-Sr2RuO4-science2014}
\bibinfo{author}{\bibfnamefont{C.~W.} \bibnamefont{Hicks}},
  \bibinfo{author}{\bibfnamefont{D.~O.} \bibnamefont{Brodsky}},
  \bibinfo{author}{\bibfnamefont{E.~A.} \bibnamefont{Yelland}},
  \bibinfo{author}{\bibfnamefont{A.~S.} \bibnamefont{Gibbs}},
  \bibinfo{author}{\bibfnamefont{J.~A.} \bibnamefont{Bruin}},
  \bibinfo{author}{\bibfnamefont{M.~E.} \bibnamefont{Barber}},
  \bibinfo{author}{\bibfnamefont{S.~D.} \bibnamefont{Edkins}},
  \bibinfo{author}{\bibfnamefont{K.}~\bibnamefont{Nishimura}},
  \bibinfo{author}{\bibfnamefont{S.}~\bibnamefont{Yonezawa}},
  \bibinfo{author}{\bibfnamefont{Y.}~\bibnamefont{Maeno}},
  \bibnamefont{et~al.}, \bibinfo{journal}{Science}
  \textbf{\bibinfo{volume}{344}}, \bibinfo{pages}{283} (\bibinfo{year}{2014}).

\bibitem[{\citenamefont{Barber et~al.}(2018)\citenamefont{Barber, Gibbs, Maeno,
  Mackenzie, and Hicks}}]{PRL2018-Sr2RuO4-Barber}
\bibinfo{author}{\bibfnamefont{M.~E.} \bibnamefont{Barber}},
  \bibinfo{author}{\bibfnamefont{A.~S.} \bibnamefont{Gibbs}},
  \bibinfo{author}{\bibfnamefont{Y.}~\bibnamefont{Maeno}},
  \bibinfo{author}{\bibfnamefont{A.~P.} \bibnamefont{Mackenzie}},
  \bibnamefont{and} \bibinfo{author}{\bibfnamefont{C.~W.} \bibnamefont{Hicks}},
  \bibinfo{journal}{Phys. Rev. Lett.} \textbf{\bibinfo{volume}{120}},
  \bibinfo{pages}{076602} (\bibinfo{year}{2018}).

\bibitem[{\citenamefont{Steppke et~al.}({2017})\citenamefont{Steppke, Zhao,
  Barber, Scaffidi, Jerzembeck, Rosner, Gibbs, Maeno, Simon, Mackenzie
  et~al.}}]{Steppke-2017Science-Sr2RuO4}
\bibinfo{author}{\bibfnamefont{A.}~\bibnamefont{Steppke}},
  \bibinfo{author}{\bibfnamefont{L.}~\bibnamefont{Zhao}},
  \bibinfo{author}{\bibfnamefont{M.~E.} \bibnamefont{Barber}},
  \bibinfo{author}{\bibfnamefont{T.}~\bibnamefont{Scaffidi}},
  \bibinfo{author}{\bibfnamefont{F.}~\bibnamefont{Jerzembeck}},
  \bibinfo{author}{\bibfnamefont{H.}~\bibnamefont{Rosner}},
  \bibinfo{author}{\bibfnamefont{A.~S.} \bibnamefont{Gibbs}},
  \bibinfo{author}{\bibfnamefont{Y.}~\bibnamefont{Maeno}},
  \bibinfo{author}{\bibfnamefont{S.~H.} \bibnamefont{Simon}},
  \bibinfo{author}{\bibfnamefont{A.~P.} \bibnamefont{Mackenzie}},
  \bibnamefont{et~al.}, \bibinfo{journal}{{Science}}
  \textbf{\bibinfo{volume}{{355}}} (\bibinfo{year}{{2017}}).

\bibitem[{\citenamefont{Kissikov et~al.}(2018)\citenamefont{Kissikov, Sarkar,
  Lawson, Bush, Timmons, Tanatar, Prozorov, Bud’ko, Canfield, Fernandes
  et~al.}}]{kissikov2018uniaxial-NC}
\bibinfo{author}{\bibfnamefont{T.}~\bibnamefont{Kissikov}},
  \bibinfo{author}{\bibfnamefont{R.}~\bibnamefont{Sarkar}},
  \bibinfo{author}{\bibfnamefont{M.}~\bibnamefont{Lawson}},
  \bibinfo{author}{\bibfnamefont{B.}~\bibnamefont{Bush}},
  \bibinfo{author}{\bibfnamefont{E.~I.} \bibnamefont{Timmons}},
  \bibinfo{author}{\bibfnamefont{M.~A.} \bibnamefont{Tanatar}},
  \bibinfo{author}{\bibfnamefont{R.}~\bibnamefont{Prozorov}},
  \bibinfo{author}{\bibfnamefont{S.}~\bibnamefont{Bud’ko}},
  \bibinfo{author}{\bibfnamefont{P.~C.} \bibnamefont{Canfield}},
  \bibinfo{author}{\bibfnamefont{R.}~\bibnamefont{Fernandes}},
  \bibnamefont{et~al.}, \bibinfo{journal}{Nat. Commun.}
  \textbf{\bibinfo{volume}{9}}, \bibinfo{pages}{1058} (\bibinfo{year}{2018}).

\bibitem[{\citenamefont{Fang and Chen}(2019)}]{fang-chen-dataset-1}
\bibinfo{author}{\bibfnamefont{Y.-W.} \bibnamefont{Fang}} \bibnamefont{and}
  \bibinfo{author}{\bibfnamefont{H.}~\bibnamefont{Chen}},
  \emph{\bibinfo{title}{{The complex non-collinear magnetic orderings in
  Ba$_2$YOsO$_6$: A new approach to tuning spin-lattice interactions and
  controlling magnetic orderings in frustrated complex oxides}}}
  (\bibinfo{year}{2019}), \bibinfo{note}{{Dataset archived in Zenodo}},
  \urlprefix\url{https://doi.org/10.5281/zenodo.3265828}.

\end{thebibliography}


\begin{thebibliography}{1}
\expandafter\ifx\csname natexlab\endcsname\relax\def\natexlab#1{#1}\fi
\expandafter\ifx\csname bibnamefont\endcsname\relax
  \def\bibnamefont#1{#1}\fi
\expandafter\ifx\csname bibfnamefont\endcsname\relax
  \def\bibfnamefont#1{#1}\fi
\expandafter\ifx\csname citenamefont\endcsname\relax
  \def\citenamefont#1{#1}\fi
\expandafter\ifx\csname url\endcsname\relax
  \def\url#1{\texttt{#1}}\fi
\expandafter\ifx\csname urlprefix\endcsname\relax\def\urlprefix{URL }\fi
\providecommand{\bibinfo}[2]{#2}
\providecommand{\eprint}[2][]{\url{#2}}

\bibitem[{\citenamefont{Treiber and
  Kemmler-Sack}(1981)}]{BYOO-expstructure-1981-german}
\bibinfo{author}{\bibfnamefont{U.}~\bibnamefont{Treiber}} \bibnamefont{and}
  \bibinfo{author}{\bibfnamefont{S.}~\bibnamefont{Kemmler-Sack}},
  \bibinfo{journal}{Z. anorg. allg. Chem. (ZAAC)}
  \textbf{\bibinfo{volume}{478}}, \bibinfo{pages}{223} (\bibinfo{year}{1981}).

\end{thebibliography}

\end{document}


\title{Supplementary Materials for: \\
  The complex non-collinear magnetic orderings in \BYOO: A new approach to tuning spin-lattice interactions and controlling magnetic orderings in frustrated complex oxides}

\author{Yue-Wen Fang$^{1,2}$}
\email{fyuewen@gmail.com}
\author{Ruihan Yang$^{2,3}$}
\author{Hanghui Chen$^{2,4}$}
\email{hanghui.chen@nyu.edu}
\affiliation{$^1$Department of Materials Science and Engineering, Kyoto University, Kyoto 606-8501, Japan \\
$^2$NYU-ECNU Institute of Physics, New York University Shanghai China \\
$^3$Department of Engineering and Computer Science, New York University Shanghai China \\
$^4$Department of Physics, New York University, New York  10003, USA
}

\date{\today}
\maketitle

\section{Details of experimental structures of B\lowercase{a}$_2$YO\lowercase{s}O$_6$}

\begin{table}[h!]
\caption{Experimental structure of ordered double perovskite oxide
  Ba$_2$YOsO$_6$ taken from Ref.~\cite{BYOO-expstructure-1981-german}.
  The space group of Ba$_2$YOsO$_6$ crystal structure is $Fm\bar{3}m$
  (No. 225).}
\begin{tabularx}{0.7\textwidth}{*{6}{Y}}
  \hline \hline 
\multicolumn{6}{c}{Cell parameters}\\
\hline
$a$  &  $b$ & $c$  &  $\alpha$  &  $\beta$ &  $\gamma$\\
                \hline
8.357~\AA &  8.357~\AA  &  8.357~\AA &  90$^{\circ}$  &  90$^{\circ}$
                                                  & 90$^{\circ}$ \\
\hline
\multicolumn{6}{c}{Atomic coordinates}\\
\hline
Site &  Element  & Wyckoff Symbol &  $X$  &  $Y$ &  $Z$ \\
\hline
Ba  &  Ba        &  $8c$    & 1/4     & 1/4   & 1/4  \\
Y    &  Y       &  $4b$   & 1/2    &  1/2   &  1/2   \\
Os   &  Os      &  $4a$    &   0     &    0     &  0     \\
O     &  O      & $24e$  &   1/4  &  0  &  0    \\
\hline
\hline
\end{tabularx}
\end{table}

\newpage
\clearpage

%

\section{Solution to Eq. (3)}
Here we show that Eq. (3) in the main text has three extremals.
\begin{equation}
\label{eq3}f(\theta, \phi) = \frac{E}{N}
=-2J_1+\alpha_1\left(\frac{13}{4}-\cos\theta+\frac{7}{4}\cos(2\theta)+2\cos\left(\frac{\theta}{2}\right)^4\cos(2\phi)\right) + E_0
\end{equation}
where $\alpha_1 = a_1 + a_2$.

Derivative of $f(\theta, \phi)$ with respect to $\theta$ and $\phi$
leads to:

\begin{eqnarray}
    \begin{cases}
\label{eq2222}
\frac{\partial f}{\partial \theta} =-4\cos^3\left(\frac{\theta}{2}\right)\cos(2\phi)\sin\left(\frac{\theta}{2}\right) + 2\Big(1-7\cos\theta\Big)\sin\left(\frac{\theta}{2}\right)\cos\left(\frac{\theta}{2}\right)=0 \\
\frac{\partial f}{\partial \phi} =\cos^4\left(\frac{\theta}{2}\right)\sin(2\phi)=0 
   \end{cases}
\end{eqnarray}

One obvious solution to Eq.~(\ref{eq2222}) is
$\cos\left(\frac{\theta}{2}\right)=0$, which leads to the collinear
antiferromagnetic ordering $E_1$ ($\theta=\pi=180^{\circ}$).

The second and third solutions to Eq.~(\ref{eq2222}) are
$\phi=\frac{\pi}{2}$, which makes $\frac{\partial f}{\partial
  \phi}=0$. Then $\frac{\partial f}{\partial \theta}$ is reduced to:

\begin{equation}
\label{eq22} \left.\frac{\partial f}{\partial
   \theta}\right\vert_{\theta=\frac{\pi}{2}}
=2\cos^2\left(\frac{\theta}{2}\right)\sin\left(\frac{\theta}{2}\right)
+(1-7\cos\theta)\sin\left(\frac{\theta}{2}\right)=0
\end{equation}
Eq.~(\ref{eq22}) has at least two solutions: one is $\theta=0^{\circ}$
(coplanar antiferromagnetic ordering $E_2$) and the other is $\theta =
\arccos\left(\frac{1}{3}\right)\simeq 71^{\circ}$ (non-collinear
non-coplanar antiferromagnetic ordering $E_3$).

The above three solutions ($E_1$, $E_2$, $E_3$) always
exist irrespective of the values of $J_1$ and $\alpha_1$.

The energy of $E_1$ is:
\begin{equation}
\label{eq23} E_1=f(\theta=\pi, \phi)= -2 J_1 + 6\alpha_1 + E_0
\end{equation}

The energy of $E_2$ is:
\begin{equation}
\label{eq24} E_2=f\left(\theta=0, \phi=\frac{\pi}{2}\right)= -2 J_1 + 2\alpha_1+E_0
\end{equation}

The energy of $E_3$ is:
\begin{equation}
\label{eq25} E_3=f\left(\theta=\arccos\left(\frac{1}{3}\right), \phi=\frac{\pi}{2}\right)=-2J_1 + \frac{2}{3} \alpha_1 +E_0
\end{equation}

\newpage
\clearpage

\section{The PBE+SOC and LDA+SOC calculations of B\lowercase{a}$_2$YO\lowercase{s}O$_6$}

\begin{figure}[h!]
	\includegraphics[angle=0,width=0.45\textwidth]{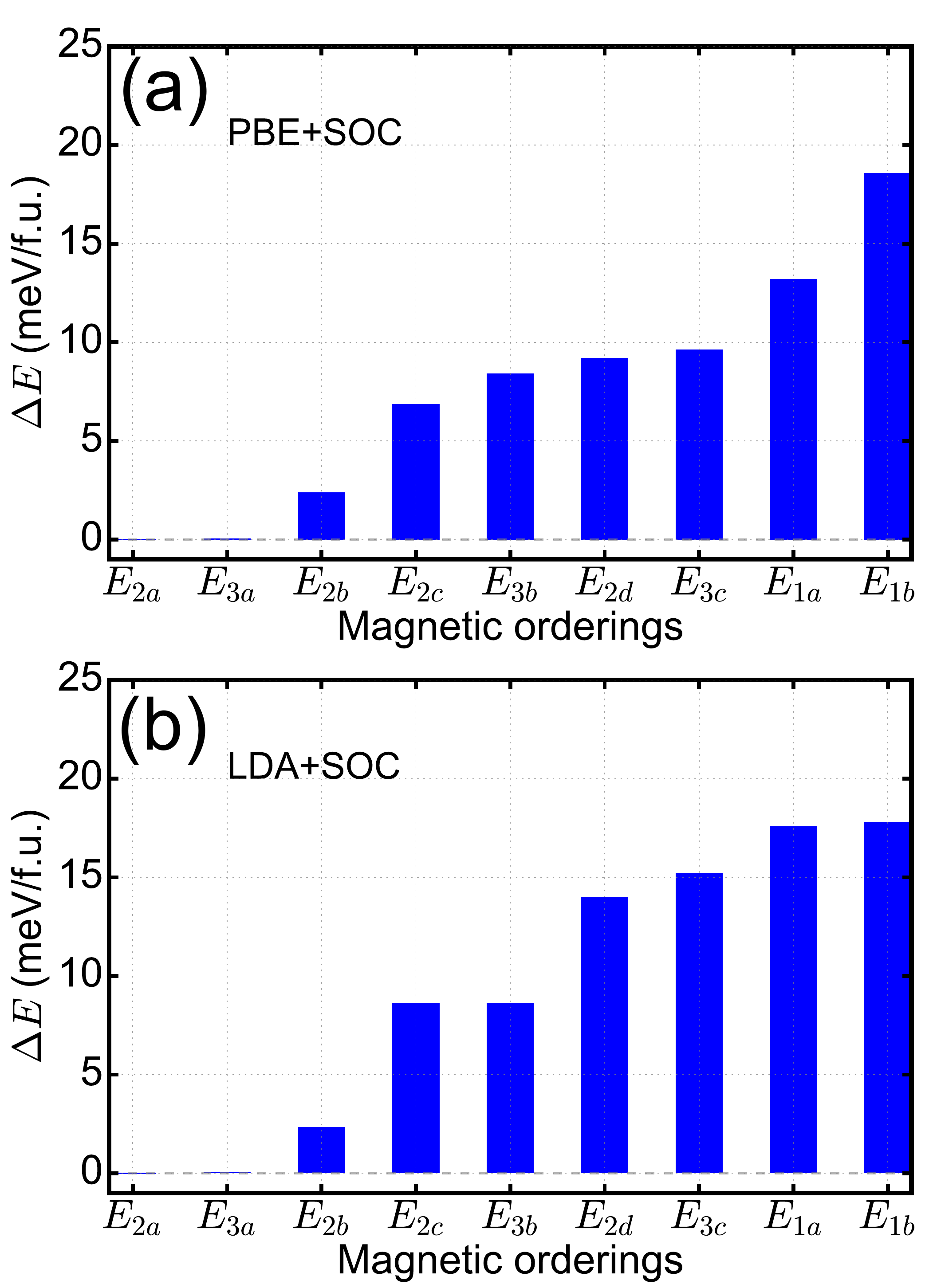}
\caption{\label{figs:LDA-PBE-SOC}
Energy sequence in ascending order of the
  nine stable antiferromagnetic orderings found in Ba$_2$YOsO$_6$ from
  (a) spin polarized DFT-PBE+SOC calculations and (b)
  spin polarized DFT-LDA+SOC calculations.
The two lowest energy states are a
  non-collinear non-coplanar state $E_{3a}$ and a coplanar state
  $E_{2a}$. The energy of $E_{2a}$ is set as the zero energy.
}
\end{figure}

In this section, we test different exchange correlation functionals in
DFT+SOC calculations to show that non-collinear antiferromagnetic
orderings are more stable than the type-I antiferromagnetic ordering
in Ba$_2$YOsO$_6$.  We re-calculate the nine long-range magnetic
orderings shown in Fig. 5 in the main text by using spin-polarized
DFT-PBE+SOC and DFT-LDA+SOC calculations. The results are shown in
Fig.~\ref{figs:LDA-PBE-SOC}.  We find that the non-collinear
non-coplanar antiferromagnetic ordering ($E_{3a}$) and the coplanar
antiferromagnetic ordering ($E_{2a}$) are almost degenerate, and they
are more stable than the type-I antiferromagnetic orderings ($E_{1a}$
and $E_{1b}$) in all the calculations.

\clearpage
\newpage


\bibliography{DP-SOC}